\documentclass[openany]{nature}
\usepackage[T1]{fontenc}
\usepackage[left]{lineno}
\usepackage{amsthm,amsmath,amssymb}
\usepackage{mathrsfs}
\usepackage{overpic}
\usepackage[T1]{fontenc}
\usepackage{multibib}
\def\emph#1 {\textit{ #1 } }
\usepackage{graphicx}
\usepackage{pdflscape}
\usepackage{hyperref}
\usepackage{threeparttable}
\usepackage{xcolor}
\usepackage{ulem}
\usepackage{cancel}
\usepackage[figuresleft]{rotating}
\usepackage{lineno}
\usepackage{caption}
\usepackage{subfigure}
\usepackage{tablefootnote}
\usepackage{amsmath, amssymb}
\usepackage{booktabs}

\makeatletter
 
\def\emph#1 {\textit{ #1 } }
\let\saved@includegraphics\includegraphics
\AtBeginDocument{\let\includegraphics\saved@includegraphics}

\makeatother

\newcommand{\apj}{Astrophys. J.}
\newcommand{\apjl}{Astrophys. J.}
\newcommand{\apjs}{Astrophys. J.}
\newcommand{\aap}{Astron. Astrophys.}
\newcommand{\mnras}{Mon. Not. R. Astron. Soc.}
\newcommand{\nat}{Nature}
\newcommand{\araa}{Ann. Rev. Astron. Astrophys.}

\newcommand{\prd}{Phys. Rev. D.}

\newcommand{\physrep}{Physics Reports}

\def\be{\begin{eqnarray}}
\def\ee{\end{eqnarray}}

\makeatletter

\def\@fnsymbol#1{\ensuremath{\ifcase#1\or \dagger\or \ddagger\or
 \mathsection\or \mathparagraph\or \|\or **\or \dagger\dagger
 \or \ddagger\ddagger \else\@ctrerr\fi}}
\makeatother

\usepackage{hyperref}
\usepackage{graphicx}
\usepackage{longtable}
\usepackage{supertabular}
\usepackage{float} 
\usepackage{authblk}

\title{Significant cocoon emission and photosphere duration stretching in GRB 211211A: a burst from a neutron star - black hole merger}
\author{Yan-Zhi Meng$^{1,2}$\thanks{E-mail: yzmeng@nju.edu.cn}, Xiangyu Ivy Wang$^{1,2}$, Zi-Ke Liu$^{1,2}$}

\begin{document}
\maketitle

\begin{affiliations}
\item School of Astronomy and Space Science, Nanjing University, Nanjing
210023, China
\item Key Laboratory of Modern Astronomy and Astrophysics (Nanjing
University), Ministry of Education, China
\end{affiliations}

\begin{abstract}
The radiation mechanism (thermal photosphere
or magnetic synchrotron)\cite{Uhm2014,Zhang2018,Burgess2020,ZhangB2020} and the progenitor\cite{Galama1998,Reeves2002,Hjorth2003,Eichler1989,Narayan1992} of gamma-ray burst (GRB) are under hot debate. Recently discovered\cite{Rast2022,Yang2022,Troja2022,Gomp2022}, the prompt long-duration ($\sim$ 10 s, normally from the collapse of massive stars\cite{Galama1998,Reeves2002,Hjorth2003}) property of GRB 211211A strongly conflicts with
its association with a kilonova (normally from the merger of two compact
objects\cite{Eichler1989,Narayan1992,Li1998,Abbott2017,Zhang2018b,Pian2017,Coulter2017}, NS-NS, NS-BH, or NS-WD, duration $\lesssim$ 2 s). In this paper, we find the probability photosphere model with a structured jet\cite{Mesz2000,Pe'er2008,Pe'er2011,Meng2018,Meng2019,Meng2021,Meng2022} can satisfactorily explain this peculiar long duration, through the duration stretching effect ($\sim$ 3 times) on the intrinsic longer ($\sim$ 3 s) duration of NS-BH (neutron star and black hole) merger\cite{Ruiz2021,Kyuto2021}, the observed empirical 2SBPL spectrum (with soft low-energy index $\alpha$ of $\sim$ -1) and its evolution\cite{Gomp2022}. Also, much evidence of the NS-BH merger origin is found, especially the well fit of the afterglow-subtracted optical-NIR light curves\cite{Rast2022} by the significant thermal cocoon emission\cite{Nakar2017,Izzo2019} and the sole thermal “red” kilonova component\cite{Metzger2017}. Finally, a convincing new explanation for the X-ray afterglow plateau\cite{Zhang2006} is revealed.
\end{abstract}



\subsection{The probability photosphere emission from the structured jet.}

\label{sec:model}

Many previous theoretical studies\cite{Pe'er2008,Pe'er2011,Belo2011} show that, in
the relativistic condition, a probability density
function $P(r,\Omega)$ should be introduced to describe the photosphere emission (see Methods). This function describes the probability for a photon
to be last scattered at any place $(r,\Omega)$ of the outflow.

The probability density function can be calculated
by (see Ref.\cite{Lund2013}) 
\begin{equation}
P(r,\Omega )=(1+\beta )D^{2}\times \frac{R_{\text{ph}}}{r^{2}}\exp \left( -%
\frac{R_{\text{ph}}}{r}\right) ,
\label{a}
\end{equation}%
where $\beta $ means the jet velocity, $D$ means the Doppler factor, and $R_{\text{ph}}$ is the photospheric radius and depends on the angular coordinate $\Omega$.

Furthermore, the GRB relativistic jet launched by the center engine will
be collimated by the surrounding material\cite{ZhaWoo2003}.
Then, the break-out jet should have angular profiles outside the isotropic core $\theta _{c}$, for both the luminosity $L$ and
Lorentz factor $\Gamma$. This is called the structured jet (see Methods). Previously, $\protect\theta _{c,\Gamma}=\protect\theta _{c,L}$ is typically assumed. Here, $\theta _{c,\Gamma}$ means the isotropic angular width for the Lorentz
factor, and $\theta_{c,L}$ means the isotropic angular width for
luminosity. 

\begin{figure*}[tbp]
\centering\includegraphics[angle=0,height=4.3in]{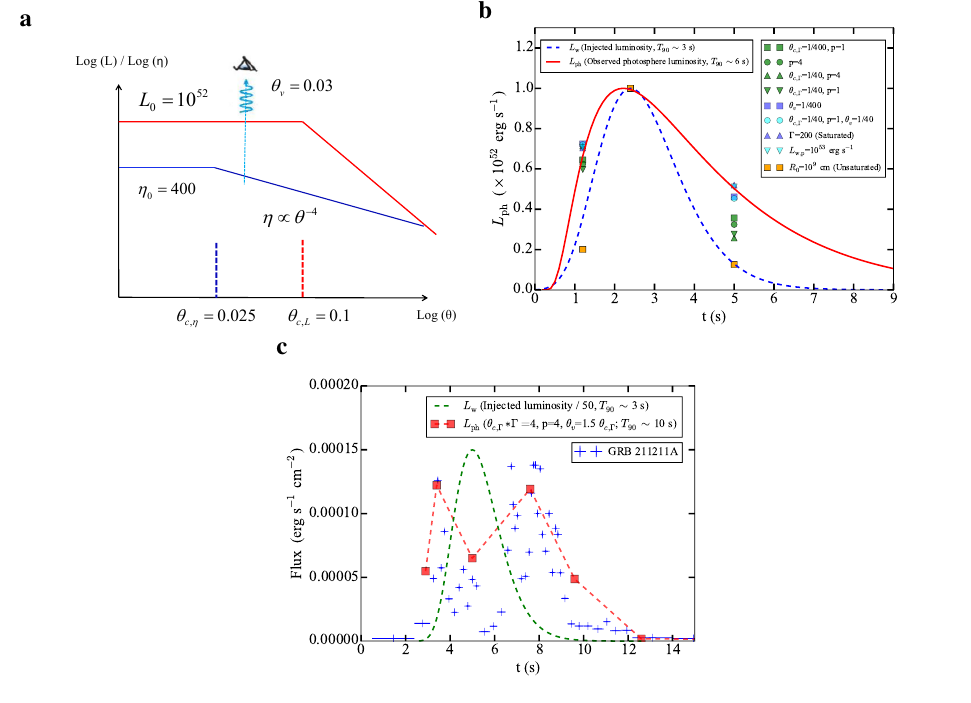} 
\caption{\textbf{The jet structure and the duration stretching effect for the saturated photosphere. a,} The considered structured jet with $\protect\theta %
_{c,\Gamma}<\protect\theta _{c,L}$ ($\Gamma$ $\simeq$ $\eta$). The considered viewing angle $\protect%
\theta_{v}$ for GRB 211211A is a bit larger than $\protect\theta_{c,\Gamma}$%
. \textbf{b,} The photosphere luminosity $L_{\text{ph}}$ profiles for different Lorentz factor structures, with the same injected duration of $T_{90}$ $\sim$ 3 s (see also Figure 16 in Ref.\cite{Meng2021}).
The observed duration can be stretched to $T_{90}$ $\sim$ 6 s, for
the strongly saturated cases (the blue and cyan markers). Note that, the $L_{\text{ph%
}}$ profiles have all been normalized to $L_{w,\text{p}}$ $=10^{52}$ erg s$%
^{-1}$ at $2.4$ s. Actually, for all the saturated cases (except for the orange box), the luminosity is 1-10 times smaller. \textbf{c,} The $L_{\text{ph}}$ profile for $\theta_{c,\Gamma} \cdot \Gamma = 4$, $\theta _{v} = 1.5 \cdot \theta_{c,\Gamma}$ (red squares). The light curve of GRB 211211A (blue plus) can be well reproduced, including the $T_{90}$ $\sim$ 10 s and the double pulses.}
\label{fig:lc}
\end{figure*}

Based on the simulation results of GRB jet\cite{ZhaWoo2003,Lazza2006,Tche2008,Geng2019,Ito2021}, we now propose that the structured jet with $\protect\theta _{c,\Gamma}<\protect\theta _{c,L}$ is likely to be commonly obtained.  (see Figure \ref{fig:lc}a). This is due to the enhanced material in the outside part, with decreasing velocity and almost constant energy (achieving the pressure equilibrium) within $\theta _{c,L}$ (see Extended Data Figure \ref{fig:sche} and further discussions in Methods). In our previous  photosphere spectrum calculations for structured jet, the constant luminosity case\cite{Meng2019} and the $\theta_{c,\Gamma}=\theta _{c,L}$ case\cite{Meng2021} are both explored. Here, we further study the $\theta _{c,\Gamma}<\theta _{c,L}$ case.

We find the luminosity structure has a negligible effect to the spectrum calculation when the viewing angle $\theta_{v}$ satisfies $\theta_{v}<\theta_{c,L}$, since relativistic emission is within $\lesssim 5/\Gamma$. Also, $\theta_{v}<\theta_{c,L}$ should be common due to the large $\theta_{c,L}$ and the decreasing luminosity outwards. In the following calculations and fitting, we adopt $\theta_{c,L}=0.1$, the typical constrained value of the jet opening angle $\theta_{\text{jet}}$. On the contrary, the Lorentz factor structure could have significant influence, since $\theta _{c,\Gamma}$ may be quite small ($\theta_{c,\Gamma} \cdot \Gamma$ $\sim$ a few) or larger $\theta_{v}$ ($\theta_{v} \gtrsim \theta_{c,\Gamma}$, our considered case for GRB 211211A) is obtained. 

The adopted Lorentz factor (or the baryon loading $\eta$, normally $\Gamma \simeq \eta$) structure is as following: 
\begin{eqnarray}
\eta (\theta ) &=&\frac{\eta _{0}}{[(\theta /\theta _{c,\Gamma
})^{2p}+1]^{1/2}}+1.2,
\label{b}
\end{eqnarray}%
where $\theta $ is the angle from the jet axis, $\eta _{0}$ is the
constant baryon loading parameter in the core with the width of $\theta
_{c,\Gamma }$, $p$ describes how the baryon loading decreases outside the
core. The constant of 1.2 donates the minimum Lorentz factor of the relativistic jet (velocity $v \sim$ 0.6 $c$, where $c$ is the speed of the light; corresponding to the highest velocity of the outside non-relativistic cocoon, see Extended Data Figure \ref{fig:sche}).

\subsection{The duration stretching effect for the saturated-acceleration photosphere.}

For the photosphere
emission, the saturated-acceleration regime is defined when $R_{\text{ph}}$ is larger than the radius $R_{s}$, where the maximum acceleration is achieved ($\Gamma=\eta$, $R_{s}=\eta \cdot R_{0}$, $R_{0}$ is the initial acceleration radius; see further discussions in Methods). The adiabatic cooling happens during $R_{s} < r \lesssim R_{\text{ph}}$, thus decreasing the observed temperature (or frequency) and luminosity.

For the injected luminosity history $L_{w}(\hat{t})$ of the center engine ($\hat{t}$ is the injection time), we use the exponential model\cite{Meng2019,Norris2005} to approximate it,  with following form:
\begin{eqnarray}
L_{w}(\hat{t} &>&\hat{t}_{s})=L_{w,p}\times \exp \left[ 2\left( \tau
_{1}/\tau _{2}\right) ^{1/2}\right]  \times \exp \left( -\frac{\tau _{1}}{\hat{t}-\hat{t%
}_{s}}-\frac{\hat{t}-\hat{t}_{s}}{\tau _{2}}\right) \text{.}
\end{eqnarray}%
where $\hat{t}_{s}$ means the start time, $\tau _{1}$ and $\tau _{2}$ are
the characteristic quantities indicating the rise and decay
timescale, respectively.  $L_{w,p}$ is the peak luminosity at $\hat{t}_{p}$, and $%
\hat{t}_{p}=\hat{t}_{s}+\left( \tau _{1}\cdot \tau _{2}\right) ^{1/2}$.

In Figure \ref{fig:lc}b, with the injected duration $T_{90}$ of $\sim$ 3 s ($\tau _{1}=32$, $\tau _{2}=0.5$, and $\hat{t}_{s}=-1.6$ are adopted; $\hat{t}_{p}=2.4$ s), we show the observed photosphere luminosity $L_{\text{ph}}$ profiles
for different Lorentz factor structures (see also Figure 16 in Ref.\cite{Meng2019}).
Obviously, for the strongly saturated-acceleration cases ($R_{\text{ph}}$ is larger, corresponding to the luminosity around the peak time), whether $\Gamma=200$ (smaller), $L_{w,\text{p}} =10^{53}$ erg s$^{-1}$ (higher) or larger $\theta_{v}$ ($\gtrsim$ $\theta _{c,\Gamma}$), the observed duration can be stretched
to $T_{90}$ $\sim$ 6 s (two times larger). While for the strongly unsaturated
case, $R_{0}=10^{9}$ cm, the observed luminosity profile is almost the same
as the injected profile, thus obtaining a similar duration. 

The reason for the above results is that, for saturated cases (actually for the higher
luminosity around the peak time), the observed luminosity around the peak time will be
significantly decreased (due to the adiabatic cooling), while the observed luminosity for earlier and later times will not suffer from this drop (with lower luminosity, $R_{\text{ph}}$ is smaller and the unsaturated condition is
obtained). In total, we discover the duration stretching effect for photosphere emission, which means that the injected luminosity duration $T_{90}$ of the center engine can be stretched to an observed duration of $\sim$ 2-3 $ T_{90}$, when $R_{s} < R_{\text{ph}}$ (saturated acceleration) is satisfied around $L_{w,\text{p}}$.

Note that, for the long GRBs, this photosphere duration stretching effect takes part for the saturated case is confirmed by comparing the duration distributions of the $\epsilon _{\gamma}\lesssim 50\%$ and $\epsilon _{\gamma }\gtrsim 50\%$ samples (see Extended Data Figure \ref{fig:dura}a, data is taken from Ref.\cite{Meng2022}).  Here, $\epsilon _{\gamma}=E_{\text{iso}}/(E_{\text{iso}} + E_{\text{k}})$, standing for the prompt efficiency. $E_{\text{iso}}$ means the isotropic energy for prompt emission, and $E_{\text{k}}$ means the energy for afterglow. Theoretically, $\epsilon _{\gamma}\lesssim 50\%$ should correspond to the saturated acceleration (smaller $\epsilon_{\gamma}$ due to the adiabatic cooling, see Methods), and its duration is indeed found 
$\sim$ 1.6 times longer than that for the $\epsilon _{\gamma }\gtrsim 50\%$ sample (with similar distribution profile).

This duration stretching effect for the saturated photosphere may
also contribute greatly to the peculiar long duration of GRB 211211A ($\sim$ 10 s) and GRB 060614 ($\sim$ 6 s). Since, according to their observed $\alpha$ - $\theta _{\text{jet}}$ distribution (see Extended Data Figure \ref{fig:alpha}a and discussions in Methods), the larger $\theta _{v}$ ($\gtrsim$ $\theta _{c,\Gamma}$; thus $\Gamma$ is smaller, see Figure \ref{fig:gamma}) is likely to be satisfied.

In Extended Data Figure \ref{fig:alpha}b, we compare the peak energy $E_{\text{p}}$ and $E_{\text{%
iso}}$ distributions of the main pulse (MP) and extended emission (EE) of GRB 211211A
and GRB 060614, with other short GRBs with EE. For these two special bursts, the $E_{\text{p}}$ is smaller of $\sim$ 2-3 times for both the MP and EE. This further supports the above saturated
scenario (due to the adiabatic cooling). The larger $\theta _{v}$ is preferred, since in
Ref.\cite{Meng2022} we find the normal short GRB sample with EE almost has $\epsilon
_{\gamma }\gtrsim 50\%$ (thus intrinsically unsaturated; with large $\theta_{c,\Gamma}$, see discussions below and Extended Data Figure \ref{fig:sche}). 

Indeed, in Figure \ref{fig:lc}c, with $\theta _{v} = 1.5 \cdot \theta_{c,\Gamma}$ and $\theta_{c,\Gamma} \cdot \Gamma = 4$, and injected duration $T_{90}$ $\sim$ 3 s, we find the light curve of GRB 211211A can be well explained by the photosphere emission, including the $T_{90}$ $\sim$ 10 s and the double pulses (stronger decrease around $L_{w,\text{p}}$).

\subsection{The photosphere explanation for the empirical 2SBPL spectrum and the spectral evolution.}

\label{sec:spec}

In Extended Data Figure \ref{fig:speccom}a, we show the theoretical photosphere spectral evolution, for larger $\theta _{c,\Gamma}$ and $\theta _{v}$ ($\theta_{c,\Gamma} \cdot \Gamma = 4$, $\theta _{v} = 1.5 \cdot \theta_{c,\Gamma}$) considered in Figure \ref{fig:lc}c. The photosphere spectra could reproduce the observed soft $\alpha$ $\sim$ -1 of GRB 211211A.

In Extended Data Figure \ref{fig:speccom}b, we show the spectral evolution of GRB 211211A for the empirical 2SBPL  (two smoothly broken power-law) model fitting, taking the best-fit parameters in Ref.\cite{Gomp2022} (accounting for the X-ray data, the 2SBPL is likely to be the better empirical model, see Methods). This evolution is quite similar to the photosphere spectral evolution in Extended Data Figure \ref{fig:speccom}a, transferring from the early 2SBPL spectrum (with high-energy exponential cutoff, which may be better, see Methods) to the late-time SBPL spectrum (due to the high-altitude effect, see Methods).

\subsection{The spectral fitting using the physical photosphere model.}

\begin{figure*}[htbp]
\vspace{-1.5cm}
\centering\includegraphics[angle=0,height=5.6in]{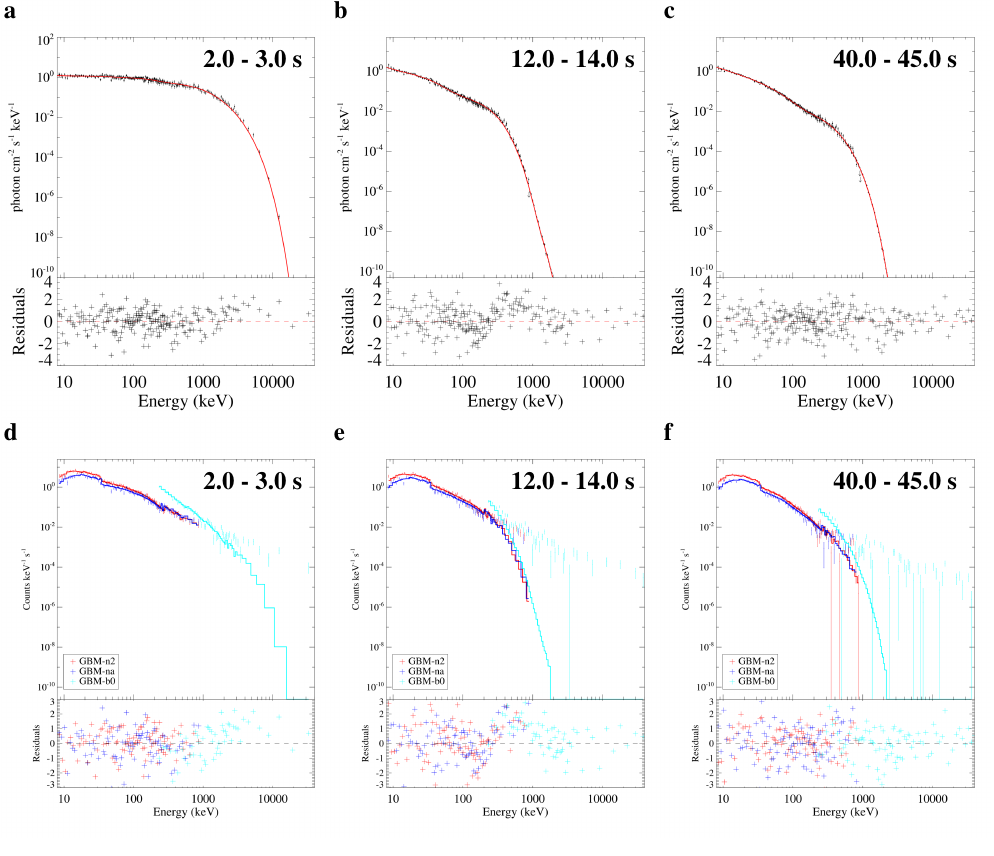} 
\caption{\textbf{The spectral fitting using the physical photosphere model, for the time-resolved spectra of GRB 211211A measured from 2.0 - 3.0 s (a and d), 12.0 - 14.0 s (b and e), 40.0 - 45.0 s (c and f). a, b and c,} De-convolved (black error bars) and modeled (red line) photon number spectra. \textbf{d, e and f,} Observed (bars) and modeled (lines) photon count spectra. }
\label{fig:specfit}
\end{figure*}

Generally, we need to convolve the model spectra with the instrumental response, the Detector Response Matrix (DRM), to compare the model spectra with the observational spectra (adopting the statistical value of BIC, Bayesian information criterion).  We use the McSpecFit package which accepts flexible user-defined spectral model\cite{Zhang2016} to perform this. The comparisons of the observed and model-convolved count spectra are shown in the bottom panels of Figure \ref{fig:specfit}, and the de-convolved observed and modeled photon number spectra are in the top panels.

The spectral fitting results of 2.0 - 3.0 s (around the peak), using the above photosphere model with structured jet (constant injected luminosity is assumed), are shown in Figure \ref{fig:specfit}, Extended Data Figure \ref{fig:par2} (parameter constraints) and Extended Data Table \ref{TABLE:MCpho} (best-fit parameters). Obviously, the photosphere model can give a rather well fit (see the residuals distribution, BIC/dof = 354/359), showing the exponential high-energy cutoff ($E_{\text{p}}$ $\sim$ 2000 keV, low-energy power-law index $\alpha_{2}$ $\sim$ - 1) combined with a smoothly broken power law in the low-energy end (the break energy $E_{\text{b}}$ $\sim$ 30 keV, low-energy power-law index $\alpha_{1}$ $\sim$ 0). 

The best-fit $\eta_{0}$ $\sim$ 700 (or $\Gamma_{0}$) is consistent with the larger $\eta$ of short GRB inferred from the optical afterglow (see Ref.\cite{Ghirlan2018}). The $\theta _{c,\Gamma}  \cdot \eta_{0}  \sim  7$ is large and $\theta _{\text{v}}  \simeq \theta _{c,\Gamma}$, just as expected above. The observed peak luminosity of 2 $\times 10^{51}$ erg s$^{-1}$ (see Ref.\cite{Yang2022}) is $\sim$ 15 times smaller than the best-fit $L_{0}$ ($10^{52.48}$ erg s$^{-1}$), due to the large $\theta_{\text{v}}$ (see Figure \ref{fig:lc}).

In addition, as shown in Figure \ref{fig:specfit}, Extended Data Figures \ref{fig:par12} and \ref{fig:par40}, and Extended Data Table \ref{TABLE:MCpho}, we give acceptable fit for the late-time spectra (12.0 - 14.0 s and 40.0 - 45.0 s) after considering reasonable injected luminosity profile. The BIC/dof = 475/358 is obtained for 12.0 - 14.0 s, and BIC/dof = 423/358 for 40.0 - 45.0 s. The best-fit structure parameters for different times are almost consistent, with large $\theta _{c,\Gamma}  \cdot \eta_{0} $ and $\theta _{\text{v}}  \simeq \theta _{c,\Gamma}$ (1.0 - 1.5 times).

\subsection{NS-BH-merger evidence and significant cocoon emission in GRB 211211A.}

\begin{figure*}[tbp]
\centering\includegraphics[angle=0,height=4.0in]{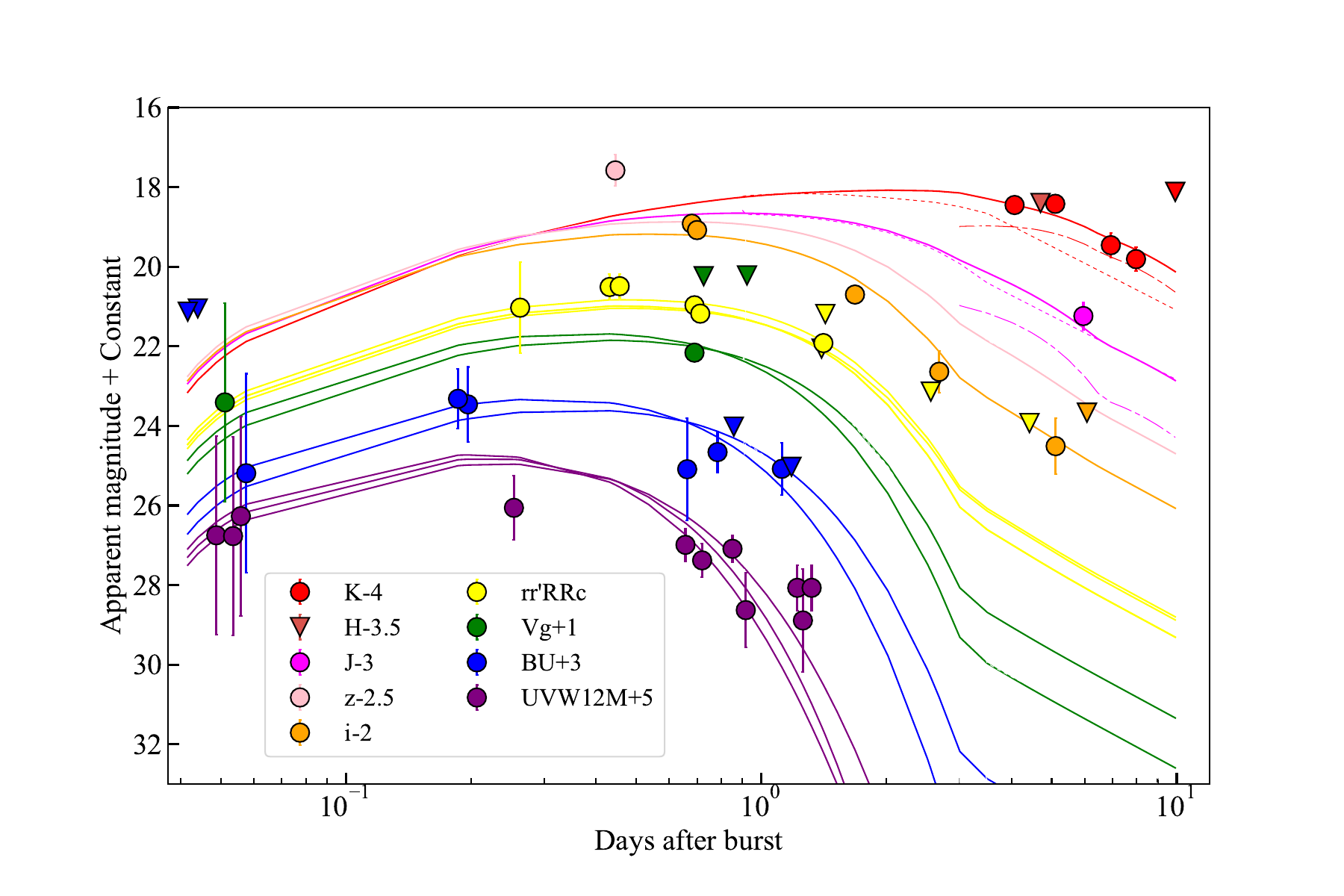} 
\caption{\textbf{Cocoon emission and NS-BH-merger evidence in GRB 211211A.} The afterglow-subtracted optical-NIR light curves of GRB 211211A (the markers)\cite{Rast2022} are well fitted by the combination (solid lines) of the early cocoon emission (dotted lines) and the late “red” kilonova component (dot-dashed lines). Except for the NIR band, the dotted lines are overlapped with the solid lines, indicating that the cocoon emission well explains the early optical light curves. Besides, the existence of a sole “red” kilonova component, without significant “blue” or “purple” kilonova components, strongly supports\cite{Metzger2017} the NS-BH merger origin of GRB 211211A (see Extended Data Figure \ref{fig:sche}).}
\label{fig:co}
\end{figure*}

\begin{figure*}[tbp]
\centering\includegraphics[angle=0,height=4.0in]{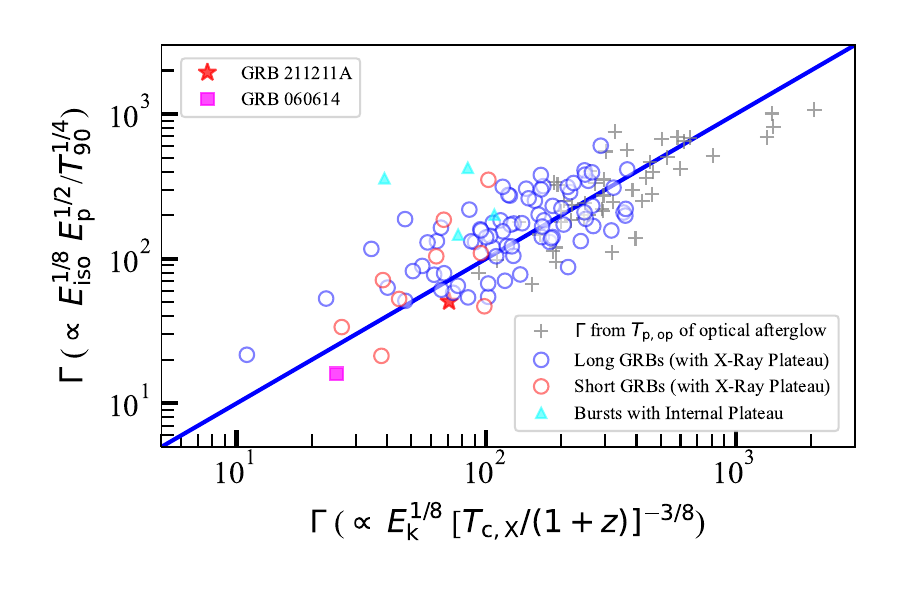} 
\caption{\textbf{Consistent estimates of $\Gamma$ from photosphere model (with the prompt emission data, $\Gamma = 17 \cdot E_{\text{iso}}^{1/8}E_{\text{p}}^{1/2}/(T_{90})^{1/4}$) and afterglow data.}  For the GRBs with X-Ray plateau (may result from large $\theta_{\text{v}}$, then smaller $\Gamma$; see Extended Data Figure \ref{fig:Xafter}), the $\Gamma$ in the light of sight may be estimated by the cutoff time of the plateau $T_{\text{c, X}}$ (taken from Ref.\cite{Tang2019}; the circles), just as the peak time of the early optical afterglow $T_{\text{p,op}}$ (the gray plus)\cite{Meng2022}. For GRB 211211A and GRB 060614, these two almost consistent estimates indicate small $\Gamma$ ($\lesssim 100$), which is compatible with the large $\theta _{\text{v}}$ consideration ($\theta _{\text{v}}  \gtrsim \theta _{c,\Gamma}$) in the above light curve and spectral analyses. The rare rate for such events within the short GRB sample with EE implies a large $\theta_{c,\Gamma}$, consistent with the expected jet structure of NS-BH merger (see Extended Data Figure \ref{fig:sche}).}
\label{fig:gamma}
\end{figure*}

According to Extended Data Figure \ref{fig:dura}b (see discussions in Methods), for the short GRBs, we propose that the $\epsilon_{\gamma}\lesssim 50\%$ sample (without EE) is likely to come from the NS-NS merger, with shorter intrinsic duration of $\sim$ 0.1 - 0.2 s\cite{Ruiz2016}. The $\epsilon_{\gamma }\gtrsim 50\%$ sample (with EE) comes from the NS-BH merger, with longer intrinsic duration of $\sim$ 0.8 s and wider duration distribution of $\sim$ 0.05 s - 3 s\cite{Ruiz2020}.

Besides, there are several other supports for this point. First, the low-energy extended emission is likely to originate from the fallback accretion of the BH, considering the comparable energy for MP and EE (especially after correcting the efficiency of EE, see Figure 5(d) in Ref.\cite{Meng2022}), and the existence of the time gap between them (disfavoring the magnetar spin-down scenario). Moreover, the luminosity of the fallback accretion from the NS-BH merger is theoretically an order of magnitude larger than that from the NS-NS merger\cite{Ross2007,Metzger2017}. Second, the short GRBs with EE typically have a small offset from their host galaxies ($\sim$ a few kpc; including GRB 211211A and GRB 060614, see Ref.\cite{Yang2022}), quite compatible with the low kick velocity for NS-BH merger\cite{Troja2008}. Third, in Extended Data Figure \ref{fig:XrayEE}, we find that all the X-ray afterglow light curves for the short GRB sample with EE show a power-law shape, while the significant plateau appears in the light curves for the sample without EE. Also, the evidence that the X-ray plateaus result from $\theta_{v}$ $\gtrsim$ $\theta_{c,\Gamma}$ is revealed below (see Extended Data Figures \ref{fig:Xafter} and \ref{fig:effi}, and Figure \ref{fig:gamma}). Then, the power-law shape and plateau well correspond to the jet structures for the NS-BH merger (larger $\theta_{c,\Gamma}$) and NS-NS merger (smaller $\theta_{c,\Gamma}$; due to much dynamical ejecta in jet propagating direction\cite{Metzger2017} and thus stronger jet-ejecta interaction) (see Extended Data Figure \ref{fig:sche}). The outliers of GRB 211211A and GRB 060614, which have EE but show plateau, can be well explained by the rare regime of $\theta_{v}$ $\gtrsim$ $\theta_{c,\Gamma}$ (with relatively large $\theta_{v}$, see Extended Data Figure \ref{fig:sche}) for NS-BH merger.

In addition, in Figure \ref{fig:co}, we find that the afterglow-subtracted optical-NIR light curves of GRB 211211A (data is taken from Tables 1 and 2 in Ref.\cite{Rast2022}) are well fitted by the combination of the dominated early cocoon emission and the late “red” kilonova component (with a mass of $M_{\text{ejr}} \simeq 0.030 M_{\odot}$), after treating more cocoon parameters than Ref.\cite{Rast2022}. The lack of significant “blue” (may have similar mass to the cocoon mass $M_{\text{eco}} \sim 0.001 M_{\odot}$) or “purple” kilonova components strongly supports the NS-BH merger origin of GRB 211211A (see Extended Data Figure \ref{fig:sche} and Methods). 

Note that we better explain the late-time i-band data, which is obviously over-estimated in Ref.\cite{Rast2022}. The parameter constraints from the fitting (see details in Methods) are shown in Extended Data Figure \ref{fig:parco}.  Also, with an energy distribution for different velocity $s = -d$ ln $E/d$ ln $v = 1$, the typical value from cocoon numerical simulation\cite{Nakar2017}, the observed evolutions of bolometric luminosity $L_{\text{bol}}$, effective temperature $T_{\text{eff}}$, and photospheric radius $R_{\text{ph}}$ (data is taken from Figure 2 in Ref.\cite{Troja2022}) can be well explained by the cocoon emission (see Extended Data Figure \ref{fig:evoco}), without the need of peculiar higher $L_{\text{bol}}$ and velocity ($\sim$ 0.6 $c$) in the early time (for the kilonova explanation)\cite{Troja2022}.

Previously, the optical cocoon signature in long GRBs was discovered\cite{Izzo2019} by analyzing the early spectra of the supernova. For short GRBs, only weak evidence (or signature) was found in the optical counterpart of GRB 170817A (Swope Supernova Survey 2017a, SSS17a; Refs\cite{Piro2018,Nicholl2021}). This is due to the intense “blue”  and “purple” kilonova emissions in GRB 170817A, produced by the NS-NS merger. Here, for GRB 211211A, which is likely to originate from the NS-BH merger, we discover the significant cocoon emission in short GRBs for the first time.

According to the above results, we consider that GRB 211211A and GRB 060614 (with EE) 
come from the NS-BH merger with intrinsic long duration ($\sim$ 3 s). In
addition, the larger $\theta_{v}$ ($\theta_v \gtrsim \theta_{c,\Gamma}$, rare event; see Extended Data Figure \ref{fig:sche}) makes them saturated (smaller $\Gamma$, see Figure \ref{fig:gamma}; $\epsilon_{\gamma}\lesssim 50\%$, see Extended Data Figure \ref{fig:effi}), resulting in the photosphere duration stretching (reaching $\sim$ 6 - 10 s duration, see Figure \ref{fig:lc}), softer $\alpha$ (see Figure \ref{fig:specfit} and Extended Data Figure \ref{fig:alpha}a), and more minor $E_{\text{p}}$ (see Extended Data Figure \ref{fig:alpha}b).

\subsection{The new explanation for the X-ray afterglow plateau: the structured jet with $\theta_{c,\Gamma}<\theta _{c,L}$ and $\theta_v > \theta_{c,\Gamma}$.} 

In Ref.\cite{Meng2022} (see Figure 7(c) there), we find $\Gamma = 17 \cdot E_{\text{iso}}^{1/8}E_{\text{p}}^{1/2}/(T_{90})^{1/4}$, derived from the photosphere model, is an excellent $\Gamma$ estimate for the bursts with the peak time of the early optical afterglow $T_{\text{p,op}}$ (also shown by the gray plus in Figure \ref{fig:gamma}). In this work, as shown in Extended Data Figure \ref{fig:Xafter} (the theoretical calculation, see Methods), we find the X-ray afterglow plateau can be explained by the structured jet with $\theta_{c,\Gamma}<\theta _{c,L}$ and $\theta_v > \theta_{c,\Gamma}$. This condition means that the $\Gamma$ in the light of sight (LOS) is much smaller, and the large cutoff time of the plateau $T_{\text{c, X}}$ should correspond to this smaller $\Gamma$ ($\Gamma  \propto E_{\text{k}}^{1/8}[T_{\text{c, X}}/(1+z)]^{-3/8}$, here $z$ is the redshift, see Ref.\cite{Ghirlan2018}).

In Figure \ref{fig:gamma}, for the large sample with plateau and $T_{\text{c, X}}$ (taken from Ref.\cite{Tang2019}), we do find consistent $\Gamma$ results from the photosphere estimate ($\Gamma = 17 \cdot E_{\text{iso}}^{1/8}E_{\text{p}}^{1/2}/(T_{90})^{1/4}$) and the afterglow estimate ($\Gamma  \propto E_{\text{k}}^{1/8}[T_{\text{c, X}}/(1+z)]^{-3/8}$), regardless of long and short GRBs (see Methods). Furthermore, the obtained $\Gamma$ is indeed much smaller (the circles; $\Gamma \lesssim 100$, with $T_{\text{c, X}} > 1000$ selection for significant plateau) than that for the bursts with early peak (the gray plus; without plateau). In Extended Data Figure \ref{fig:effi}, for this sample with significant plateau, $E_{\text{iso}}/E_{\text{k}} \lesssim 1$ is found, indicating a small prompt efficiency $\epsilon _{\gamma }\lesssim 50\%$ (resulting from the smaller $\Gamma$ and the photosphere model, $\epsilon _{\gamma } \propto \Gamma^{8/3}$).  These consistent $\Gamma$ estimates, the smaller $\Gamma$, and the lower efficiency strongly support the above explanation for the X-ray afterglow plateau.

Both the X-ray afterglow plateau exists in GRB 211211A and GRB 060614 (see Ref.\cite{Yang2022}). As shown in Figure \ref{fig:gamma}, above two $\Gamma$ estimates obtain consistently smaller $\Gamma$ for these two bursts ($\Gamma \simeq 70$ for GRB 211211A; quite close to the constrained $\Gamma = 73$ from afterglow modeling in Ref.\cite{Rast2022}, see Table 3 there). This further supports the $\theta_{c,\Gamma}<\theta _{c,L}$ and $\theta_v \gtrsim \theta_{c,\Gamma}$  consideration above (the off-core case may be named), from the light curve (Figure \ref{fig:lc}), the spectral fitting (Figure \ref{fig:specfit}) and the NS-BH merger origin (Figure \ref{fig:co} and Extended Data Figure \ref{fig:sche}; larger $\theta_{c,\Gamma}$ and rare rate for $\theta_v \gtrsim \theta_{c,\Gamma}$).

In physics research, a pure blackbody is generally expected for thermal emission. But, recent studies (especially this work) for the GRB spectrum reveal that, in the relativistic condition, a pure spherical shell photosphere should change to
a probability photosphere, obtaining a multi-color blackbody ($\alpha \sim 0$). This relativistic probability photosphere should be treated in many other astrophysics and physics regions. Also, the GRB photosphere emission is normally powered by the cooling of the high-temperature accretion disk, through  the neutrino annihilation (for neutrino-dominated accretion flow,
namely NDAF\cite{Poph1999}). Thus, our study (achieving the parameters from the fitting) provides a new and vital path to studying the accretion disk.

The relativistic jet widely exists in astrophysics objects (AGN, TDE, high-mass X-ray binaries), whose structure contains rich physics information and is crucial to the observed characteristics. Recent studies (especially this work) indicate that $\theta _{c,\Gamma}<\theta _{c,L}$ is likely to exist commonly, with which the typical GRB prompt emission ($\alpha \sim -1$) and afterglow properties (the X-ray plateau, see Extended Data Figure \ref{fig:Xafter} and Methods) can both be reproduced. Our findings should provide an important base for the X-ray and gamma-ray polarization studies since the polarization degree strongly depends on the jet structure. In addition, the structured jet should be accompanied by the outward non-relativistic cocoon. In GRB 211211A, we reveal the significant optical cocoon emission, for the first time in short GRBs.

The gravitational wave from the NS-BH coalescence has been detected (see Methods). And it is long believed that, the short GRBs should consist of two subsamples, originating respectively from the NS-NS merger and NS-BH merger. Here, based on the duration distribution (wider range of $\sim$ 0.05 s - 3 s, and a longer typical value of $\sim$ 0.4 s), the sore “red” kilonova component and other arguments, we consider that the short GRB sample with extended emission (along with GRB 211211A and GRB 060614) originates from the NS-BH merger. Our claimed distinguished duration distributions and kilonova property could be further tested for a larger sample in the future. Meanwhile, future gravitational wave observation can also check our opinion.

\subsection{Data Availability} The {\it Fermi}/GBM data are publicly available at \url{https://heasarc.gsfc.nasa.gov/W3Browse/fermi/fermigbrst.html}. The Konus-Wind data are publicly available at \url{https://vizier.cds.unistra.fr/viz-bin/VizieR?-source=J/ApJ/850/161}. The Swift data are publicly available at \url{https://www.swift.ac.uk/archive/ql.php}.

\subsection{Code Availability} Upon reasonable request, the codes (mostly in Mathematical and Python) used to produce the results and figures will be provided.    \\  
\\

\begin{methods}

\subsection{The gravitational wave observation for the NS-BH coalescence.}

The NS-BH coalescence has been confirmed recently by gravitational wave
observation (GW200105 and GW200115), in LIGO's and Virgo's third observing
runs\cite{Abbott2021}. For GW200105, the masses of the BH and NS are 8.9$_{-1.5}^{+1.2}$ $M_{\odot }$ and
1.9$_{-0.2}^{+0.3}$ $M_{\odot }$, respectively. While for GW200115, the
masses are 5.7$_{-2.1}^{+1.8}$ $M_{\odot }$ and 1.5$_{-0.3}^{+0.7}$ $M_{\odot }$,
respectively. Unfortunately, no electromagnetic counterpart (GRB or
kilonova) has been identified. The reason may be the low projected-aligned
spin for the BH, based on GW observations. Namely, these two mergers are
plunging events\cite{Zhu2021,Hu2022}. However, the BHs in some observed high-mass X-ray binaries, which could be the progenitors of the NS-BH system, are claimed to have a high
spin\cite{McClin2011}. So far, whether the
GRB can be produced by NS-BH coalescence is still uncertain. More detected
NS-BH coalescences in the future will surely settle down this. In this
work, for GRB 211211A, we find convincing evidence that it (and other
short GRBs with EE) comes from the NS-BH merger.

\subsection{The kilonova thermal emissions from the NS-NS merger and NS-BH merger.}

The neutron-rich ejecta, from the NS-NS merger or NS-BH merger,
will undergo rapid neutron capture (r-process\cite{Latt1976}) nucleosynthesis, producing
the heaviest elements like platinum and Lanthanide. Then, the radioactive
decay of these unstable nuclei will power a rapidly evolving ($\sim$ days), supernova-like transient in NIR or optical band, known as “kilonova”\cite{Li1998,Metzger2017}.

For the NS-NS merger, the kilonova emission consists of the “red” component (the opacity $\kappa=10 $ cm$^{2}$ g$^{-1}$, from lanthanide-rich dynamical ejecta in the equatorial plane), the “blue” component ($\kappa=0.5 $ cm$^{2}$ g$^{-1}$, from lanthanide-free dynamical ejecta in the polar directions, produced by shock heating in the contact interface between the merging stars; the dynamical shock enriches the electron fraction and thus lowers the neutron fraction), and the “purple” component ($\kappa \sim 3 $ cm$^{2}$ g$^{-1}$, from intermediate-opacity isotropic disk wind ejecta; moderate neutrino irradiation from the longer-lived NS remnant lowers the lanthanide fraction) (see Ref.\cite{Metzger2017} and Extended Data Figure \ref{fig:sche}). 

For the NS-BH merger, the kilonova emission only contains the “red” components
from the dynamical ejecta (in the equatorial plane, by tidal forces that disrupt the NS) and the disk wind ejecta.

\subsection{The cocoon thermal emission.}

When the jet propagates crosses the stellar envelope (for long GRBs) or the dynamical ejecta (for short GRBs), the dense material in the envelope or ejecta will be spilled sideways, forming a cocoon\cite{Nakar2017} that engulfs the jet and collimates it to make the structured jet. The GRB jet dissipates significant (may comparable) energy of itself to the cocoon through the shock. The adiabatic cooling of shock-heated material (outside the structured jet, see Extended Data Figure \ref{fig:sche}) produces a transient quite similar to the blue kilonova (at optical band, peaks at $\sim$ 1 day or much earlier).

\subsection{Cocoon plus “red” kilonova model fitting for the afterglow-subtracted optical-NIR data\cite{Rast2022}.}

We fit the afterglow-subtracted photometry with the modified model (cocoon plus “red” kilonova) using the MOSFiT (Modular Open Source Fitter for Transients\cite{Guil2018}) platform. The fitting goodness for EMCEE sampling is given by the WAIC (“Watanabe-Akaike information criterion”\cite{Sumio2010}). Here, WAIC $\sim$ 21 is obtained for 49 data points. For the nested sampling with DYNESTY, the total model evidence is ln(Z) $\sim$ 21, almost the same as that obtained in Ref.\cite{Rast2022} (ln(Z) $\sim$ 20-25).  The shown results are for the EMCEE sampling.

The mass of the “red” kilonova component, $M_{\text{ejr}} \simeq 0.030 M_{\odot}$, is consistent with that in Ref.\cite{Rast2022} ($M_{\text{ejr}} \simeq 0.025 M_{\odot}$). For DYNESTY sampling, higher $M_{\text{ejr}}$ is preferred ($M_{\text{ejr}} \simeq 0.05 M_{\odot}$), $s$ may also be higher. A significant improvement for our fitting, using the cocoon component instead of the “blue” and “purple” kilonova components\cite{Rast2022}, is the better explanation of the late-time (1.68 days, 2.68 days and 5.11 days) i-band data. 

\subsection{The photosphere thermal model for $\gamma$-ray prompt emission.}

The existence of the photospheric
emission is the basic prediction of the classical fireball model\cite{Good1986,Pacz1986} for GRB, because the optical depth $\tau $ at the jet base is much larger than unity\cite{Pi1999}.
As the fireball expands and the optical depth drops down, the
internally trapped thermal photons finally escape at the photosphere ($\tau
=1$). Indeed, based on the spectral analysis, a quasi-thermal component has
been found in several \textit{Fermi} GRBs (especially in GRB 090902B\cite{Abdo2009}). But whether the typical observed Band function\cite{Band1993} (smoothly joint broken power law) or cutoff power law (CPL) can be explained by the photosphere emission, namely the
photosphere emission model, remains unknown. If this scenario is true, the
quasi-thermal spectrum should be broadened. Theoretically, two different
broadening mechanisms have been proposed: subphotospheric dissipation
(namely the dissipative photosphere model\cite{Pe2004,Gian2007,Belo2010}) or geometric broadening (namely the probability photosphere model\cite{Pe'er2008,Pe'er2011,Lund2013,Deng2014,Meng2018,Meng2019,Meng2021}).

\subsection{Previous observed supports for the photosphere emission model.}

First, a quasi-thermal component has been found in a great amount of BATSE
GRBs\cite{Ryde2009} and several \textit{Fermi} GRBs
(especially in GRB 090902B and GRB 170817A)\cite{Abdo2009,Zhang2018,Zhang2018b}.
Second, lots of bursts have a low-energy spectral
index $\alpha $ harder than the death line (or the maximum value, $\alpha$ = 
$-$2/3) of the basic synchrotron model, especially for the short GRBs and
the peak-flux spectrum\cite{Kan2006,ZhaBB2011}. Third, the spectral width is found to be quite narrow for a significant fraction of GRBs\cite{AxBo2015}.
Fourth, for a half or more of the GRBs, the cutoff power law is the best-fit
empirical model, indicating that the photosphere emission model can
naturally interpret their high-energy spectrum. Fifth, recently in Ref.\cite{Meng2022}, by
dividing the GRB sample into three sub-samples based on the prompt
efficiency $\epsilon_{\gamma}$, the observed $E_{\text{p}}$ - $E_{\text{iso}}$ distribution\cite{Amati2002} of each sub-sample can be perfectly explained by the photosphere emission model.
Also, for each subsample, the characteristics of the X-ray afterglow and
optical afterglow are well consistent with the predictions of the photosphere
emission model.

\subsection{The probability photosphere model.}

For the traditional photosphere model, the photosphere emission is all
emitted at the photospheric radius $R_{\text{ph}}$, where the optical depth
for a photon propagating towards the observer is equal to unity ($\tau =1$).
But, if only there is an electron at any position, the photon should have a
probability to be scattered there. For an expanding fireball, the photons
can be last scattered at any place in the fireball with a certain
probability. Thus, the traditional spherical shell photosphere is changed to
a probability photosphere, namely the probability photosphere model\cite{Pe'er2008}. 

For the probability photosphere model, the observed photosphere spectrum is
the overlapping of a series of blackbodies with different temperatures, thus
its low-energy spectrum is broadened. After considering the jet with angular
structure, the observed typical low-energy photon index $\alpha \sim -1.0$\cite{Kan2006,ZhaBB2011},
spectral evolution and $E_{\text{p}}$ evolutions (hard-to-soft evolution or $E_{\text{p}}$%
-intensity tracking\cite{Liang1996}) can be reproduced\cite{Lund2013,Meng2019,Meng2021}, from the theoretical perspective.

\subsection{The structured jet with $\protect\theta _{c,\Gamma}<\protect%
\theta _{c,L}$.}

For GRBs, the jet launched by the center engine will
be collimated, by the gas envelope of the progenitor star for long GRBs %
\cite{ZhaWoo2003}, and the dynamical ejecta for short GRBs\cite{Aloy2005,Ross2013}.
Thus, the structured jet\cite{Dai2001,Rossi2002} should both exist. Noteworthily, based
on the unusual performance of the prompt emission and the afterglow of GRB
170817A (the first joint detection of short GRB and gravitational wave), a
structured jet in it is strongly favored\cite{Lyman2018,Ghirlanda2019}. 

Normally, for a structured jet, $\theta _{c,\Gamma }=\theta _{c,L}$ is assumed for simplicity. But whether this is the real
situation is quite uncertain and without proper reasons. Considering that $\theta _{c,\Gamma }=\theta _{c,L}$ is true, the material
density within (or beyond) the core should be rather isotropic. This may conflict
with the collimation condition. Theoretically, the progenitor envelope (or
dynamical ejecta) is matter-dominated and should make the shocked jet have
an enhanced material density at larger angle when collimation happens.
Then, because $\Gamma (\theta )\propto L/M(\theta )$, the Lorentz factor
should start to decrease even when the $L$ remains constant (see Figure 3 in %
Ref.\cite{Lazza2006} and similar discussion in Ref.\cite{Meng2021}). Besides, $\theta
_{c,\Gamma }<\theta _{c,L}$ is supported in many simulations, including both
hydrodynamical ones\cite{ZhaWoo2003,Lazza2006,Ito2021} and
magnetohydrodynamical ones (MHD, Refs.\cite{Tche2008,Geng2019}).

On the other hand, as shown in Extended Data Figure \ref{fig:sche}, the jet energy $E=\Gamma M c^{2}$ (mainly for the inner isotropic core, $M_{\text{jet}} \sim 10^{-6}-10^{-5} M_{\odot}$\cite{Ghirlan2018}, and $\Gamma \sim 100-1000$) is comparable to the outside cocoon energy $E \simeq M v^{2}$ ($M_{\text{cocoon}} \sim 10^{-3}-10^{-1} M_{\odot}$, and $v \sim 0.1-0.6 c$; Ref.\cite{Nakar2017}). Thus, the large $\theta_{c,L}$, may extending to the non-relativistic cocoon region, is reasonable (may available for most bursts, and at least for a part). Besides, numerical simulations for the jet-cocoon interaction\cite{Nakar2017} suggest roughly constant energy within the cocoon (including the jet region, especially the mildly relativistic region; see Figure 2 therein), namely $dE/dv \propto v^{-s} =v^{-1}$.

\subsection{Saturated and unsaturated acceleration, $\epsilon _{\gamma }\lesssim 50\%$ and $\epsilon _{\gamma }\gtrsim 50\%$\cite{Meng2022} respectively.}

For the photosphere emission model, $\epsilon _{\gamma }\lesssim 50\%$ should correspond to the saturated
acceleration ($R_{\text{ph}}\gtrsim R_{s}$; $\Gamma =\eta $) case, since $E_{\text{iso}}/E_{\text{k}}=[\eta Mc^{2}(R_{%
\text{ph}}/R_{s})^{-2/3}]/\Gamma Mc^{2}=(R_{\text{ph}}/R_{s})^{-2/3}$ $%
\lesssim 1$. Here, $R_{\text{ph}}\propto L_{\text{iso}}/\Gamma ^{3}$. The adiabatic cooling exists at $r \gtrsim R_{s}$. 

While, $\epsilon _{\gamma }\gtrsim 50\%$ should correspond to the unsaturated acceleration ($R_{\text{ph}}\lesssim R_{s}$; $\Gamma \lesssim \eta $) case, since $E_{\text{iso}}/E_{\text{k}}=\eta Mc^{2}/\Gamma Mc^{2}=\eta /\Gamma $ $\gtrsim 1$. The adiabatic cooling does not exist.

\subsection{The duration distribution test for long and short GRBs.}

As mentioned above, the photosphere duration stretching expects that the bursts with 
$\epsilon _{\gamma }\lesssim 50\%$ (saturated) should have a longer duration
than that with $\epsilon _{\gamma }\gtrsim 50\%$. In Extended Data Figure \ref{fig:dura}a, we compare the duration distributions of the $\epsilon _{\gamma}\lesssim 50\%$ and $\epsilon _{\gamma }\gtrsim 50\%$ samples (see Ref.\cite{Meng2022}) for
the long GRBs. As expected, the $\epsilon _{\gamma }\lesssim 50\%$ sample has
a much longer ($\sim$ 0.2 dex, namely $\sim$1.6 times) duration. Also, apart
from the mean values, these distributions are similar.

Surprisingly, for the short GRBs (see Extended Data Figure \ref{fig:dura}b), the $\epsilon _{\gamma }\lesssim 50\%$ sample has a much shorter ($\sim$ 0.1 dex, namely $\sim$ 1.3 times) duration.
Furthermore, the distribution of the $\epsilon _{\gamma }\lesssim 50\%$ sample is less extended. Considering these two properties, we think these
two distinguished samples may originate from different sources. According to
the simulations of NS-NS\cite{Ruiz2016} and NS-BH mergers\cite{Kyuto2021,Ruiz2020,Ruiz2021}, the short GRBs from NS-NS merger typically have a smaller duration of $\sim$ 0.1 - 0.2 s\cite{Ruiz2016}, entirely
consistent with the result of the $\epsilon_{\gamma }\lesssim 50\%$ sample
(Note that the mean value $\sim$ 10$^{-0.5}$ should be decreased by 1.6 - 2 times
due to the photosphere duration stretching effect). While, the short GRBs from the NS-BH merger
typically have a larger duration of $\sim$ 0.8 s\cite{Ruiz2020}, more consistent with the
result of the $\epsilon _{\gamma }\gtrsim 50\%$ sample (for the MP).

Besides, for the NS-BH merger, since the disk mass $M_{\text{disk}}$ (crucial
to the duration) depends on many parameters of the NS and BH, as follows
(see Ref.\cite{Ruiz2021}):
\begin{eqnarray}
\frac{M_{\text{disk}}}{M_{\text{NS}}}\approx 0.415\cdot R^{1/3}(1-2C)-0.148%
\frac{R_{\text{ISCO}}}{R_{\text{NS}}}.
\label{c}
\end{eqnarray}%
Here, $M_{\text{NS}}$ and $R_{\text{NS}}$ are the mass and radius of the NS, 
$C=M_{\text{NS}}/R_{\text{NS}}$ means the NS compactions. $R=M_{\text{BH}%
}/M_{\text{NS}}$ and $M_{\text{BH}}$ is the mass of the BH. And, $R_{\text{%
ISCO}}$ is the radius of the innermost stable circular orbit, which strongly
depends on the BH spin $a_{\text{BH}}$. The duration distribution for the NS-BH merger is expected to be more
extended, just similar to the distribution of the $\epsilon _{\gamma
}\gtrsim 50\%$ sample (could be smaller to $\sim$ 0.05 s, and larger to $\sim$
3 s). 

\subsection{The theoretically predicted $\protect\alpha$ - $\protect\theta %
_{c,\Gamma}$ distribution and observed $\protect\alpha$ - $\protect\theta %
_{\text{jet}}$ distribution.}

As mentioned above, the observed overlapped photosphere spectrum should be
broadened. Previous studies\cite{Lund2013,Meng2019,Meng2021} show that, for a uniform jet $\alpha$ can be softened to $\alpha$ $\sim$ 0. Whereas, for the structured jet considered here, $%
\alpha$ could be much softer, reproducing the observed typical $\alpha$ $\sim$ -1.
Because that, according to Equation $(\ref{a})$, the extra low-energy contribution
within angle of $\sim$ $5/\Gamma$ is proportional to $D^{2}=[\Gamma (\theta
)\cdot (1-\beta (\theta )\cdot \cos \theta_{_{\text{LOS}}})]^{-2}$ ($%
\theta_{_{\text{LOS}}}$ is the angle aparted from the line of sight). When
smaller $\Gamma$ is obtained for $\theta_{_{\text{LOS}}}> 0$, the extra
low-energy component will be greatly enhanced.

In Extended Data Figure \ref{fig:alpha}a, we show the schematically theoretically-predicted $%
\alpha$ - $\theta _{c,\Gamma}$ distribution (the dashed lines)\cite{Lund2013,Meng2019}. Two categories
are obviously seen: smaller $\theta _{c,\Gamma}$ and
larger $\theta _{c,\Gamma}$. Notice that, acturally $\theta _{c,\Gamma
}<\theta _{c,L}$ $\lesssim$ $\theta _{\text{jet}}$ is likely expected for structured
jet. But, we think these three quantities are likely to change with the same
trend. Namely, with larger $\theta _{\text{jet}}$, $\theta _{c,\Gamma }$ and $\theta
_{c,L}$ are both larger.  For smaller $\theta _{c,\Gamma}$ ($\theta%
_{c,\Gamma} \cdot \Gamma$ $\sim$ 1), regardless of the $\theta _{v}$, the decreasing-$\Gamma$ component always exists within $\theta_{_{\text{LOS}}}$ $\lesssim$ $5/\Gamma$. Thus, the $\alpha$ is quite soft, clustered around $\sim$ - 1.
For larger $\theta _{c,\Gamma}$, $\theta _{v}$ has great effect. Normally, $%
\theta _{v} < \theta _{c,\Gamma}$ is expected. Then, within $\theta_{_{%
\text{LOS}}}$ $\lesssim$ $5/\Gamma$, the isotropic-$\Gamma$ component
dominates and the decreasing-$\Gamma$ component contributes less. Thus, the $%
\alpha$ should be quite hard, ranging from 0 to $\sim$ - 0.7. However, when $%
\theta _{v} \gtrsim \theta _{c,\Gamma}$ is obtained with lower chance, the
decreasing-$\Gamma$ component dominates, $\alpha$ can range from $\sim$ - 1 to
- 2.  
          
Interestingly, the above theoretically predicted $\alpha$ - $\theta
_{c,\Gamma}$ distribution seems to be consistent with the observed $\alpha$ - $%
\theta_{\text{jet}}$ distribution ($\theta_{\text{jet}}$ is taken from Ref.\cite{Du2021}), shown also in Extended Data Figure \ref{fig:alpha}a. For smaller $\theta _{\text{jet}}$ (may $\theta _{c,\Gamma}$ also), the $\alpha$ is clustered around $\sim$ - 1. When $\theta _{\text{jet}}$ is
larger, the $\alpha$ becomes much harder (reaching $\sim$ 0), except for a few
bursts (including GRB 211211A and GRB 060614). These outliers are likely to
be due to the larger $\theta _{v}$ ($\gtrsim$ $\theta _{c,\Gamma}$).

\subsection{The photosphere explanation for the $\alpha$-flux-tracking
evolution\cite{Li2019,Yang2022}.}

As shown in Extended Data Figure \ref{fig:ee}b, the softer $\alpha$ can be obtained in the earlier time (0.3 s). Unlike the origin of the softer $\alpha$ in the later time (due to the delayed low-energy high-altitude emission emitted around the peak time, see discussions in Refs.\cite{Meng2019, Li2019}), it is due to the comparability of the peak energies for the isotropic-$\Gamma$ component and the decreasing-$\Gamma$ component. These two peak energies both strongly depend on the luminosity and have distinguished dependences (since $\Gamma$ and $\theta _{v}$ are different) if any one is in the saturated-acceleration regime. For the strongly saturated (at the peak time) case, such as our adopted parameters and GRB 211211A, in the earlier time (0.3 s) the peak energy of the decreasing-$\Gamma$ component is smaller (but not extremely, $\sim$ 10 times) than that of the isotropic-$\Gamma$ component, thus contributing a great part to the low-energy spectrum and making the $\alpha$ softer. So, combined with the softer $\alpha$ after the peak time, the special $\alpha$-flux-tracking evolution\cite{Li2019,Yang2022} in a few bursts (including GRB 211211A) is explained.

\subsection{The 2SBPL spectra in GRB 211211A.}

For the spectra of GRB 211211A, just as stated in Ref.\cite{Gomp2022} (with low-energy BAT data), the 2SBPL model is likely to be the better empirical model   (statistic/dof = 480/353 for 2SBPL model, statistic/dof = 642/355 for Band function, of 3.0 $\pm $ 1.0 s). In Ref.\cite{Yang2022}, similar conclusion is drawn. For the fit with the CPL model, the residuals show obvious rising trend in the low-energy ($\sim$ 8 - 30 keV) end, indicating a much harder ($\alpha$ $\sim$ - 0.5  to 0) component there (pgstat/dof = 440.78/362, of 3.4 - 3.5 s). The low-energy hard component may exist in all the bursts (with almost fixed energy break $E_{\text{b}}$ $\sim$ 20 keV, as discussed in the following for the photosphere model). But considering the high redshift (z $\sim$ 1 - 2) for most bursts, the energy break will decrease to $\sim$ 8 keV (beyond the \textit{Fermi} energy range, thus not obvious). For GRB 211211A, with z = 0.076, the low-energy hard component is significant.

\subsection{The drawbacks for the synchrotron origin\cite{Gomp2022} of the 2SBPL spectrum.}

 First, as shown in Figure 3 and Table 1 of Ref.\cite{Gomp2022}, $\alpha_{1}$ is all harder than the synchrotron death line $\alpha$ = $-$ 2/3. Second, the high-energy index $\beta$ is almost close to $\sim$ - 3, indicating a cutoff in the high-energy end. Third, $\alpha$  $\sim$ - 1 according to Ref.\cite{Yang2022}, while $\alpha_{2}$ $\sim$ - 3/2 is obtained in Ref.\cite{Gomp2022}. This conflict is likely to come from the inaccurate power-law formular in the high-energy end for the 2SBPL model (thus the statistic of the CPL model is comparable to that of the 2SBPL model, even though its low-energy fitting is bad). Since for the CPL spectrum, towards the higher energy, the $\alpha$ will gradually change from -1 to - Inf, mimicking a much softer $\alpha$ $\sim$ - 3/2 when the power-law formular is taken. In total, the best-fit empirical model for GRB 211211A is likely to be the CPL high-energy spectrum combined with a smoothly broken power law in the low-energy end, which resembles the photosphere spectrum in Extended Data Figure \ref{fig:speccom}a.

\subsection{The low-energy break $E_{\text{b}}$ explanation by the photosphere model.}

Notice that, for the photosphere model, the low-energy break is naturally expected. Since the low-energy blackbody component comes from much earlier injection (with larger $\theta_{_{\text{LOS}}}$, then arriving later and obtaining lower energy due to the Doppler effect). In real GRB and our calculation, limited injection duration ($\sim$ a few seconds) is expected. So, the lowest blackbody energy (mainly depends on the injection duration) should correspond to the injection of $\sim$ seconds earlier, and regardless of the prompt injection. Then, an almost fixed low-energy break is predicted (for $\sim$ 1 second, $E_{\text{b}}$ $\sim$ 10 keV, see the red and green lines of Figure 2 in Ref.\cite{Meng2019}). The photosphere low-energy break ($\sim$ 10 keV) has been shown in Figures 6 - 9 of Ref.\cite{Meng2019} (especially the red lines). For GRB 211211A, as shown in Figure 3 and Table 1 of Ref.\cite{Gomp2022}, the almost fixed low-energy break ($\sim$ 20 - 30 keV) indeed exists, strongly favoring the  photosphere origin.

\subsection{The possible photosphere explanation for the extended emission lasting $\sim$ 100 s.}

As mentioned above, the observed late-time (a few tenths seconds) low-energy ($\sim$ keV)  SBPL spectrum could be produced by the photosphere high-altitude effect (see Extended Data Figure \ref{fig:speccom}a). But, whether the theoretical flux decay can match the flux decay of observed EE is very crucial.  In Extended Data Figure \ref{fig:ee}a, we compare the theoretical flux decay with that of GRB 211211A. Obviously, the observed decay could be reproduced. Notice that, an extra injection at $\sim$ 20 - 30 s is considered, which may come from the fallback accretion of the black hole, for the NS-BH merger. The photosphere high-altitude effect extends this $\sim$ 10 s duration to $\sim$ 100 s.

\subsection{The smoother decay of the X-ray afterglow for the structure jet with $\theta_{c,\Gamma }<\theta _{c,L}$.} 

As shown in Refs.\cite{Dai2001,Rossi2002}, for the structured jet with off-axis $\theta _v$ ($\theta_v > \theta_{c,L}$, and $\theta_{c,\Gamma }=\theta _{c,L}$), the observed smoother component (close to a plateau, flux $f$ $\sim$ $t_{\text{obs}}^{-0.5}$) before normal decay ($f$ $\sim $ $t_{\text{obs}}^{-1}$) in most X-ray afterglows\cite{Zhang2006} can be explained. However, the off-axis condition should be unusual, which conflicts with the usual appearance of the smoother decay. 

In this work, as shown in Extended Data Figure \ref{fig:Xafter}, we find the usual on-axis condition ($\theta_v <\theta _{c,L}$,  see Figure \ref{fig:lc}a) can also reproduce the observed smoother decay, when $\theta_{c,\Gamma}<\theta _{c,L}$ is adopted. This is quite reasonable, since the peak time of the afterglow $T_{p}$ strongly depends on the $\Gamma$ ($\Gamma \propto (E_{\text{k}}^{{}})^{1/8}\cdot \lbrack T_{p}^{{}}/(1+z)]^{-3/8}$)\cite{Ghirlan2018}. The significant $\Gamma$ structure within $\theta _{c,L}$ will surely result in the overlapping of the afterglow components with different arriving times.

For $\theta_v \gtrsim \theta_{c,\Gamma}$ considered here, the $\Gamma$ in the light of sight is smaller ($T_{p}$ is larger), contributing to the normal decay at the later time. The inner jet (with larger $\Gamma$, larger $\theta_{_{\text{LOS}}}$) will contribute to the earlier component. The flux will be suppressed due to the larger $\theta_{_{\text{LOS}}}$, causing the observed smoother decay (or the plateau of GRB 211211A and GRB 060614, see Ref.\cite{Yang2022}).

\subsection{Consistent $\Gamma$ estimates from $\Gamma = 17 \cdot E_{\text{iso}}^{1/8}E_{\text{p}}^{1/2}/(T_{90})^{1/4}$ and $\Gamma  \propto E_{\text{k}}^{1/8}[T_{\text{c, X}}/(1+z)]^{-3/8}$.} 

For the estimate from $T_{\text{c, X}}$ in Figure \ref{fig:gamma}, a basic constant of $1.27 \times [17 \times 9 \times 9/(2^{10} \times 3.14 \times 4)]^{1/8}$ is adopted, as for the $T_{\text{p,op}}$ in Ref.\cite{Meng2022} (the gray plus; homogeneous medium with density $n=1$  cm$^{-3}$ and $E_{\text{k}}= 5 \cdot E_{\text{iso}}$ are still assumed). But another factor of $\sim$ 2.0 is needed for the long GRBs to match these two equations completely. We think it is due to the slight difference between the defined $T_{\text{c, X}}$ and the real peak time for the observed smaller $\Gamma$ ($T_{\text{c, X}}$ may be $\sim$ 4.0 times larger), since the turning after the plateau is quite smooth. The real peak time may be the earliest turning time ($T_{\text{c, X}}$ donates the middle), and the smooth turning component may originate from the weak emission of the most outside jet region (with the smallest $\Gamma$, arriving later). Other scenarios, such as smaller density, wind medium or different constant of the photosphere equation, for this off-core (large $\theta _{\text{v}}$) case may also be possible.

For the short GRBs (the constant of the photosphere equation should be $\sim$ 17/1.8), adopting the above factor of $\sim$ 2.0, an extra constant of $\sim$ 2.5 is needed. We think it is due to the smaller density $n \sim 2.5^{-8} \sim 0.001$ cm$^{-3}$ (indeed found for the short GRBs\cite{Berger2014,Connor2020}) or the smaller photosphere constant than 17/1.8.

Obviously, the bursts with internal plateau (all are long GRBs, considering the factor of $\sim$ 2.0) significantly deviate from the other samples. This indicates that they do have a different origin, likely from the late-time injection of the magnetar\cite{Dai1998,Zhang2001}.
     
\end{methods}

\begin{addendum}
 \item We thank Bin-Bin Zhang for kindly providing the spectral fitting tools and Jin-Jun Geng for calculating the afterglow. We also thank Xue-Feng Wu and Jun Yang for helpful discussions. Y.-Z.M. is supported by the National Postdoctoral Program for Innovative Talents (grant no. BX20200164). We acknowledge the support by the National Key Research and Development Programs of China (2018YFA0404204, 2022YFF0711404, 2022SKA0130102), the National Natural Science Foundation of China (Grant Nos. 11833003, U2038105, 12121003), the science research grants from the China Manned Space Project with NO.CMS-CSST-2021-B11, and the Program for Innovative Talents.   We also acknowledge the use of public data from the Fermi Science Support Center, the Swift and the Konus-Wind.

 \item[Author Contributions] YZM initiated the study and proposed the scientific opinions. YZM and XIW fit the data. YZM, XIW, and ZKL wrote the paper.

 \item[Author Information] Correspondence and requests for materials should be addressed to 
Y.-Z.M (yzmeng@nju.edu.cn).

\item[Competing Interests] The authors declare that they have no competing interests.

\end{addendum}

\setcounter{table}{0}

\setcounter{figure}{0}

\makeatletter
\renewcommand{\figurename}{Extended Data Fig.} \renewcommand{%
\tablename}{Extended Data Table.}

\makeatother

\captionsetup[table]{name={\bf Extended Data Table}}
\captionsetup[figure]{name={\bf Extended Data Figure}}
\begin{figure*}[tbp]
\centering\includegraphics[angle=0,height=4.0in]{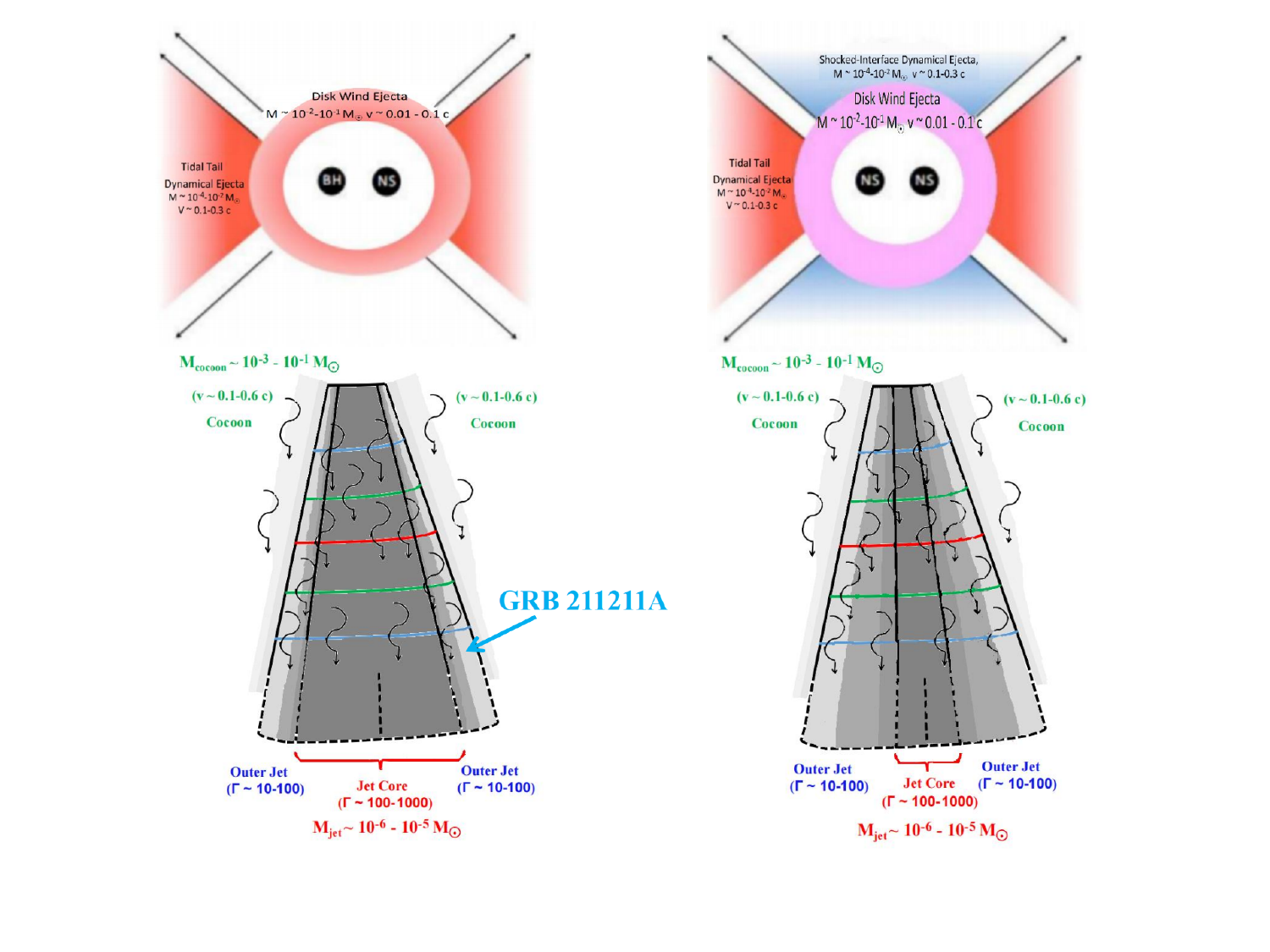}
\caption{\textbf{The schematic diagram of distinguished kilonova components and jet structures for the NS-BH and NS-NS mergers.} For the NS-NS merger, the kilonova emission\cite{Metzger2017} consists of the “red” component (from lanthanide-rich dynamical ejecta in the equatorial plane), the “blue” component (from lanthanide-free dynamical ejecta in the polar directions, produced by shock heating), and the “purple” component (from intermediate-opacity isotropic disk wind ejecta). For the NS-BH merger, the kilonova emission only contains the “red” components from the dynamical ejecta (in the equatorial plane) and the disk wind ejecta.  As much dynamical ejecta exists in the jet propagating direction, for the NS-NS merger, the structured jet (mainly the outer mild-relativistic part) should be more significant, namely with a smaller isotropic core $\theta_{c,\Gamma}$. For the NS-BH merger, $\theta_{c,\Gamma}$ should be larger. The rare rate of $\theta_{v}$ $\gtrsim$ $\theta_{c,\Gamma}$ is consistent with that for GRB 211211A and GRB 060614. Moreover, the large $\theta_{v}$ naturally explains the observations of kilonova and cocoon emissions (see Figure \ref{fig:co}).}
\label{fig:sche}
\end{figure*}

\begin{table}
\caption{\textbf{Best-fit parameters (see the meanings in Extended Data Figure \ref{fig:parco}) using cocoon plus “red” kilonova model for the afterglow-subtracted optical-NIR light curves in GRB 211211A}}
\label{TABLE:MCkilo}\center%
\begin{tabular}{lccl}
\toprule Parameters & Value \\
\midrule 
\textbf{“red” kilonova:} &  \\
$M_{\text{ejr}}$ ($M_{\odot}$) & $0.030_{-0.006}^{+0.008}$ \\
$v_{\text{ejr}}$ (km s$^{-1}$) & $96270.712_{-17457.589}^{+16709.240}$  \\
\midrule
\textbf{Cocoon:} &  \\
$M_{\text{eco}}$ ($M_{\odot}$) & $0.001_{-0.000}^{+0.000}$ \\
$v_{\text{ejco}}$ (km s$^{-1}$) & $74187.427_{-6701.947}^{+6812.600}$  \\
$\cos \theta_{\text{cocoon}}$  & $0.641_{-0.160}^{+0.140}$ \\
$s$  &  $-0.111_{-0.127}^{+0.116}$  \\
$t_{\text{shock}}$ (s) & $37.026_{-9.182}^{+9.182}$ \\
$T_{\text{shock,cool}}$ (K) & $2302.477_{-251.847}^{+192.993}$  \\
\midrule
$t_{\text{exp}}$ (days) & $-0.014_{-0.006}^{+0.005}$ \\
\bottomrule 
\end{tabular}%
\end{table}

\begin{figure*}[tbp]
\centering\includegraphics[angle=0,height=2.33in]{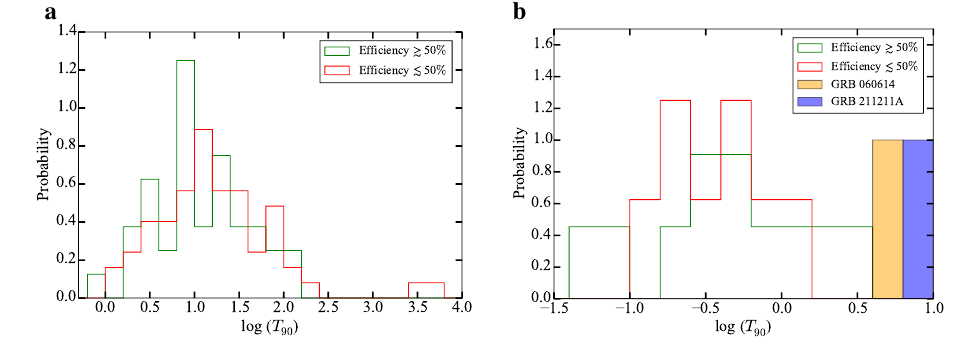} 
\caption{\textbf{The duration distributions of the $\protect\epsilon_{\protect\gamma %
}\lesssim 50\%$ (saturated, red) and $\protect\epsilon_{\protect\gamma %
}\gtrsim 50\%$ (unsaturated, green) samples, for long and short GRBs. a,} The
duration distribution for long GRBs. As expected by the
photosphere duration stretching effect, the $\protect\epsilon _{\protect\gamma }\lesssim 50\%$
sample has a much longer duration ($\sim$ 1.6 times). \textbf{b,} The duration distribution for short
GRBs, including GRB 211211A and GRB 060614. Considering the longer duration ($\sim$ 0.8 s)
and more extended distribution ($\sim$ 0.05 - 3.0 s), the $\protect\epsilon_{\protect\gamma %
}\gtrsim 50\%$ sample (possessing EE; along with GRB 211211A and GRB 060614)
may originate from the NS-BH merger\cite{Ruiz2020,Ruiz2021}. The even longer duration of GRB 211211A
and GRB 060614 is likely to be caused by the larger $\protect\theta_{v}$
(transforming to the saturated regime, and thus $\protect\epsilon_{\protect%
\gamma }\lesssim 50\%$).}
\label{fig:dura}
\end{figure*}

\begin{figure*}[tbp]
\centering\includegraphics[angle=0,height=2.23in]{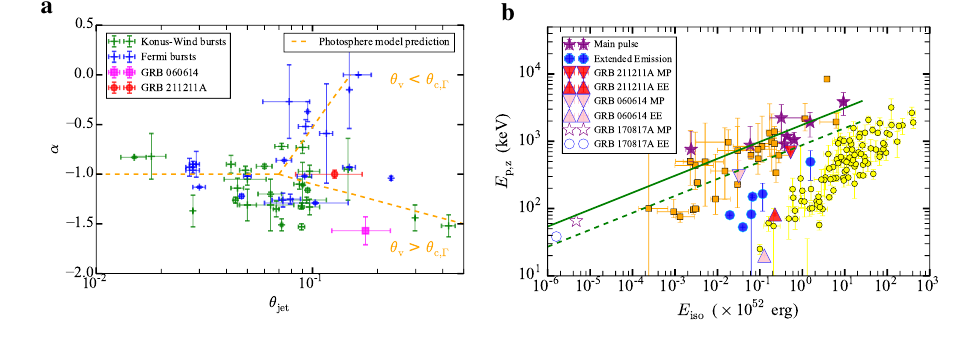} 
\caption{\textbf{The saturated-acceleration supports from the $\protect\alpha$ - $\protect\theta _{\text{jet}}$ and $E_{\text{p}}$ - $E_{\text{iso}}$ distributions, for GRB 211211A and GRB 060614. a,} The consistence of the theoretically predicted $\protect\alpha$ - $%
\protect\theta _{c,\Gamma}$ distribution (the orange dashed lines) and
observed $\protect\alpha$ - $\protect\theta _{\text{jet}}$ distribution ($\protect%
\alpha$ for the Konus-Wind bursts, marked by the green plus; $\protect\alpha$
for the \textit{Fermi} bursts, marked by the blue plus, $\protect\theta_{\text{jet}}$ is
mainly taken from Ref.\cite{Du2021}). The distributions for GRB 211211A (red circle) and
GRB 060614 (magenta square) both possess larger $\protect\theta _{\text{jet}}$ and
softer $\protect\alpha$, indicating of the larger viewing angle $\protect%
\theta_{v}$ ($\protect\theta_{v}$ $\gtrsim$ $\protect\theta_{c,\Gamma}$). \textbf{b,} The
comparison of the $E_{\text{p}}$ - $E_{\text{iso}}$ distributions of the MP
and EE for GRB 211211A and GRB 060614, with other short GRBs possessing EE.
The smaller $E_{\text{p}}$ can be seen, indicating of the saturated
condition (with adiabatic cooling; likely due to larger $\protect\theta_{v}$).}
\label{fig:alpha}
\end{figure*}

\begin{figure*}[tbp]
\centering\includegraphics[angle=0,height=2.3in]{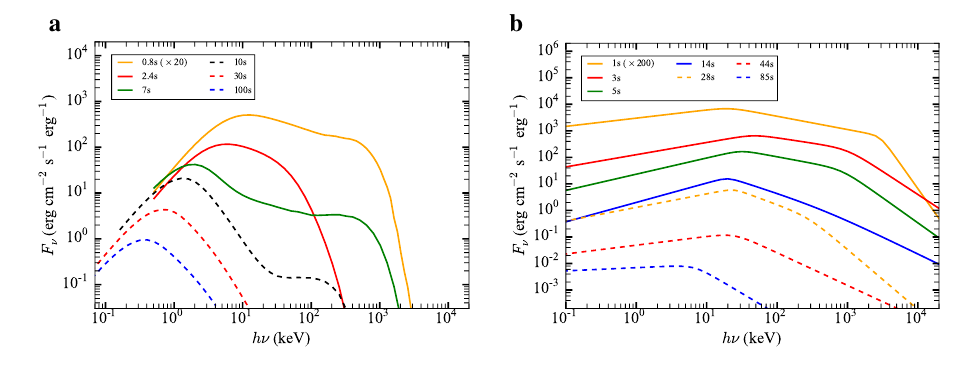} 
\caption{\textbf{The comparison of the spectral evolutions of the theoretical photosphere model and GRB 211211A (for the empirical 2SBPL model fitting\cite{Gomp2022}). a,} The spectral evolution of theoretical photosphere model ($\theta_{c,\Gamma} \cdot \eta = 4$, $\theta _{v} = 1.5 \cdot \theta_{c,\Gamma}$; and $\tau _{1}=32$, $\tau _{2}=0.5$, $\hat{t}_{s}=-1.6$). The spectrum transfers from the early 2SBPL spectrum (with high-energy exponential cutoff) to the late-time SBPL spectrum. \textbf{b,} The spectral evolution of GRB 211211A.}
\label{fig:speccom}
\end{figure*}

\begin{figure*}[tbp]
\centering\includegraphics[angle=0,height=2.3in]{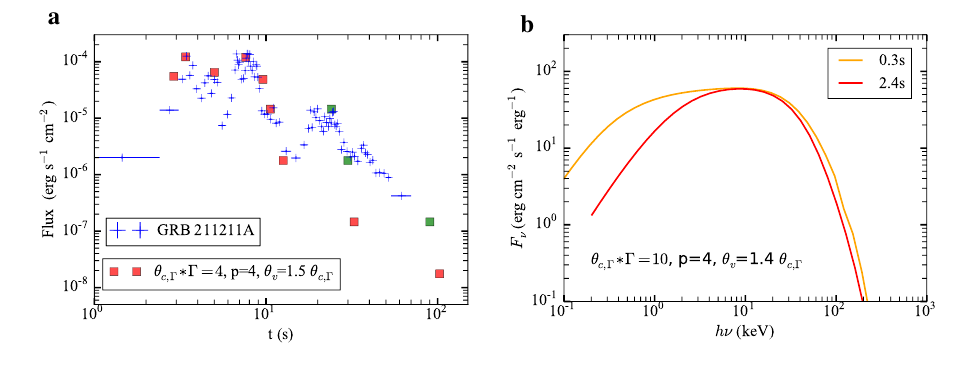} 
\caption{\textbf{The photosphere explanations for the extended emission (lasting $\sim$ 100 s) and the $\alpha$-flux-tracking
evolution\cite{Li2019}. a,} Comparison of the late-time light curve (extended emission, blue plus) of GRB 211211A and the theoretical photosphere flux decay (red and green squares). The green squares consider an extra injection at $\sim$ 20 - 30 s, which may come from the fallback accretion of the black hole, for the NS-BH merger. \textbf{b,} Comparison of the theoretical photosphere spectra for the peak time (2.4 s, red line, without the double pulses for $\theta_{c,\Gamma} \cdot \Gamma_{0} = 10$, $\theta _{v} = 1.4 \cdot \theta_{c,\Gamma}$) and the earlier time (0.3 s, orange line).  The softer $\alpha$ can be obtained in the earlier time (0.3 s). }
\label{fig:ee}
\end{figure*}

\begin{figure*}[tbp]
\centering\includegraphics[angle=0,height=6.0in]{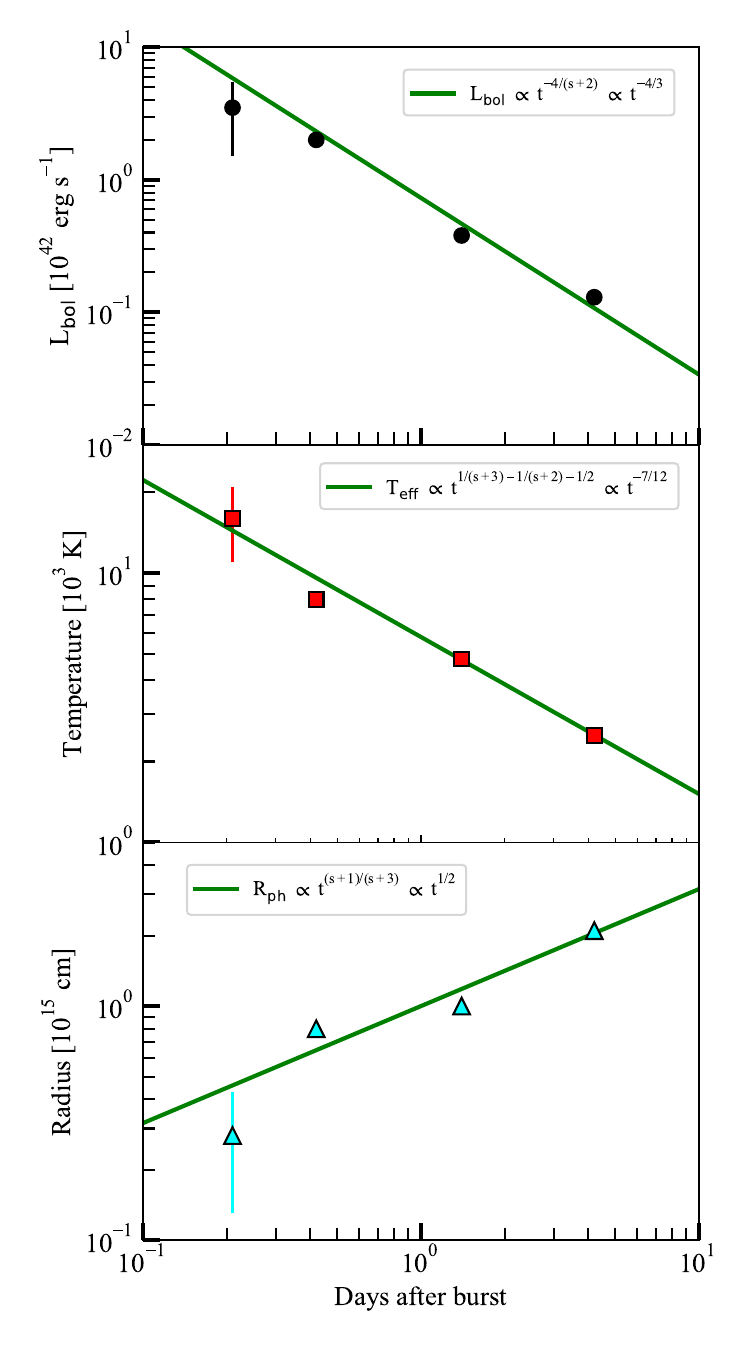} 
\caption{\textbf{The cocoon explanation of the evolutions of bolometric luminosity, effective temperature, and photospheric radius for the afterglow-subtracted optical-NIR data of GRB 211211A.} The data points (markers) are taken from Ref.\cite{Troja2022}. The solid lines represent the theoretical power-law evolutions of cocoon emission\cite{Piro2018}, for $s = -d$ ln $E/d$ ln $v = 1$ (the typical value from cocoon numerical simulation\cite{Metzger2017}). Obviously, the observed evolutions can be well reproduced by the cocoon emission.}
\label{fig:evoco}
\end{figure*}

\begin{figure*}[tbp]
\centering\includegraphics[angle=0,height=2.2in]{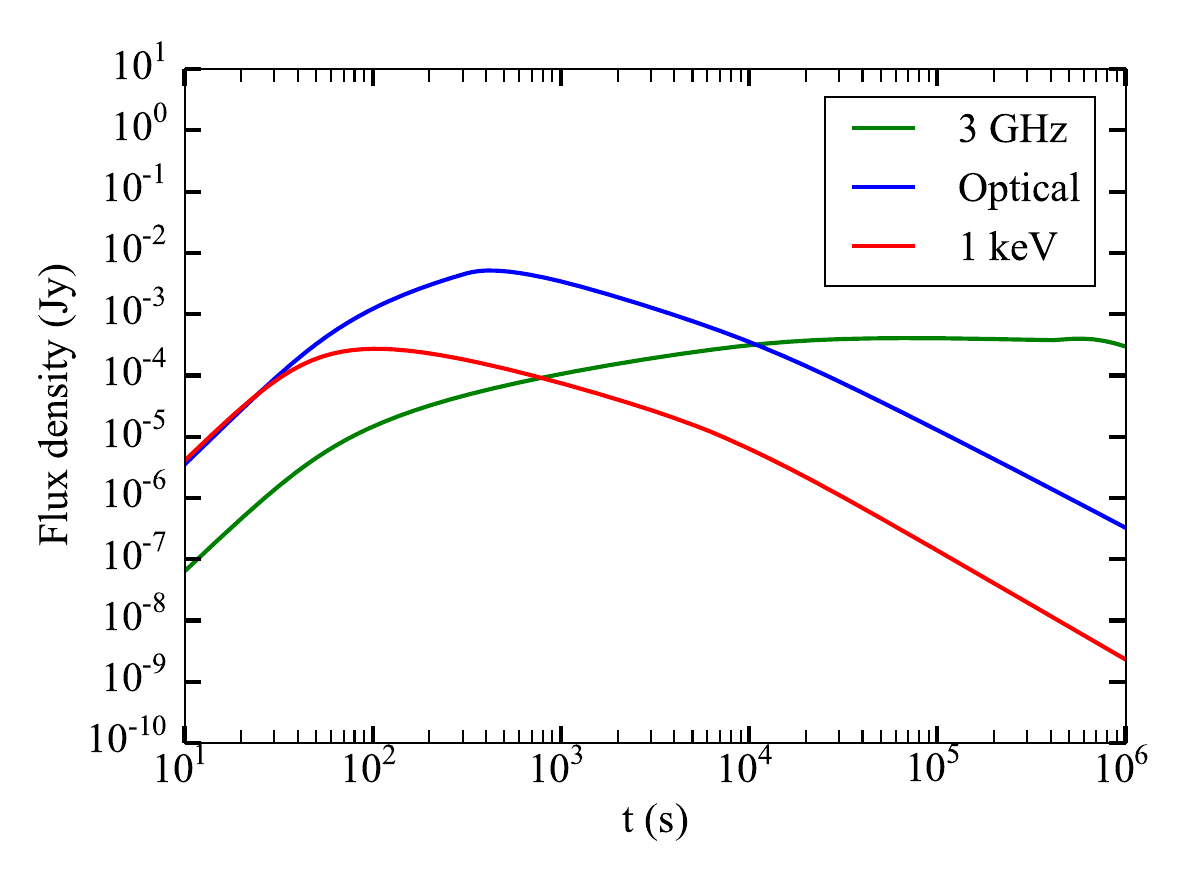} 
\caption{\textbf{The smoother decay (before normal decay) of the X-ray afterglow for the structured jet with $\theta_{c,\Gamma }<\theta _{c,L}$ and $\theta _{v}$ $\gtrsim$ $\theta _{c,\Gamma}$.} The adopted parameters are $\Gamma_{0}= 400$, $\theta_{c,\Gamma}= 0.02$, $\theta_{c,L}= 0.08$, $\theta_{v}= 0.03$, $\theta_{\text{jet}}= 0.15$. }
\label{fig:Xafter}
\end{figure*}

\begin{figure*}[tbp]
\centering\includegraphics[angle=0,height=2.1in]{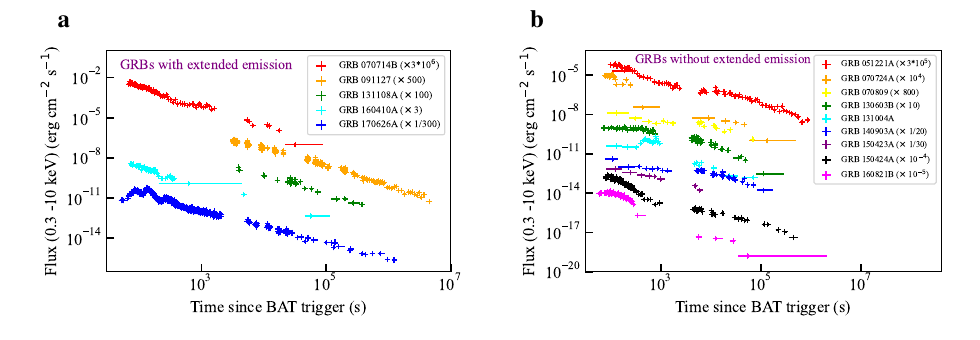} 
\caption{\textbf{Comparison of the X-ray afterglow light curves for the short GRB samples with (left) and without (right) extended emission. a,} The power-law shapes of the X-ray afterglow light curves for the short GRB sample with EE. \textbf{b,} The X-ray afterglow light curves show a significant plateau for the short GRB sample without EE. If the plateaus do result from $\theta_{v}$ $\gtrsim$ $\theta_{c,\Gamma}$ (see Extended Data Figure \ref{fig:Xafter}), the power-law shape and plateau, for bursts with and without EE, well correspond to the jet structures for the NS-BH merger (larger $\theta_{c,\Gamma}$) and NS-NS merger (smaller $\theta_{c,\Gamma}$) (see Extended Data Figure \ref{fig:sche}).}
\label{fig:XrayEE}
\end{figure*}

\begin{figure*}[tbp]
\centering\includegraphics[angle=0,height=2.2in]{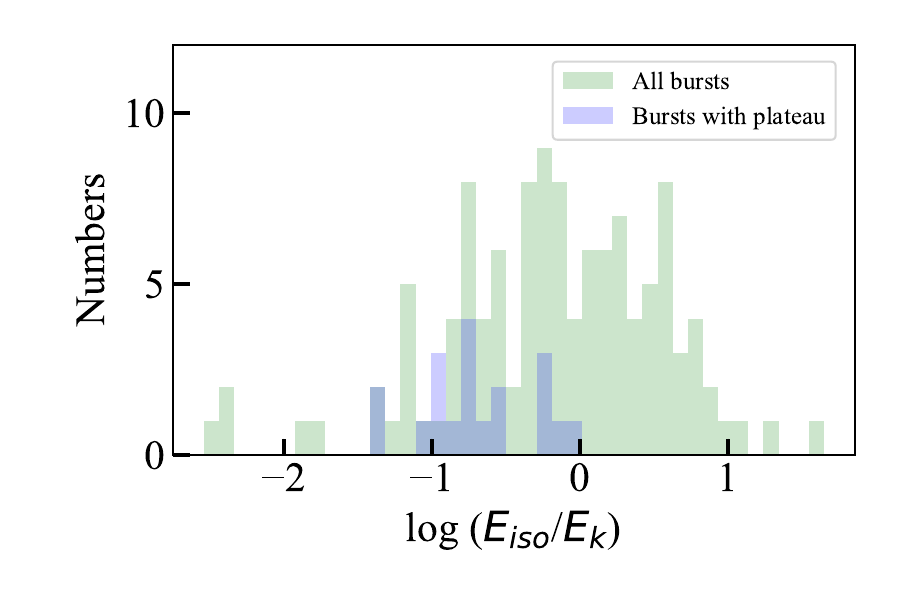} 
\caption{\textbf{Comparison of the $E_{\text{iso}}/E_{\text{k}}$ distributions for the sample with significant plateau (purple) and the whole sample (green).} The distribution for the whole sample is taken from Figure 10 in Ref.\cite{Meng2022}. And the sample with plateau is taken from Ref ($T_{\text{c, X}} \gtrsim 1000$ is adopted, to omit the weak plateau). Obviously, $E_{\text{iso}}/E_{\text{k}} \lesssim 1$ is obtained for the sample with significant plateau, indicating a small prompt efficiency ($\epsilon _{\gamma }\lesssim 50\%$). This is quite consistent with the smaller $\Gamma$ (see Figure \ref{fig:gamma}) and larger $\theta_{v}$ (see Extended Data Figure \ref{fig:Xafter}), under the photosphere framework for prompt emission ($\epsilon _{\gamma } \propto \Gamma^{8/3}$).
}
\label{fig:effi}
\end{figure*}

\begin{figure*}[tbp]
\centering\includegraphics[angle=0,height=6.2in]{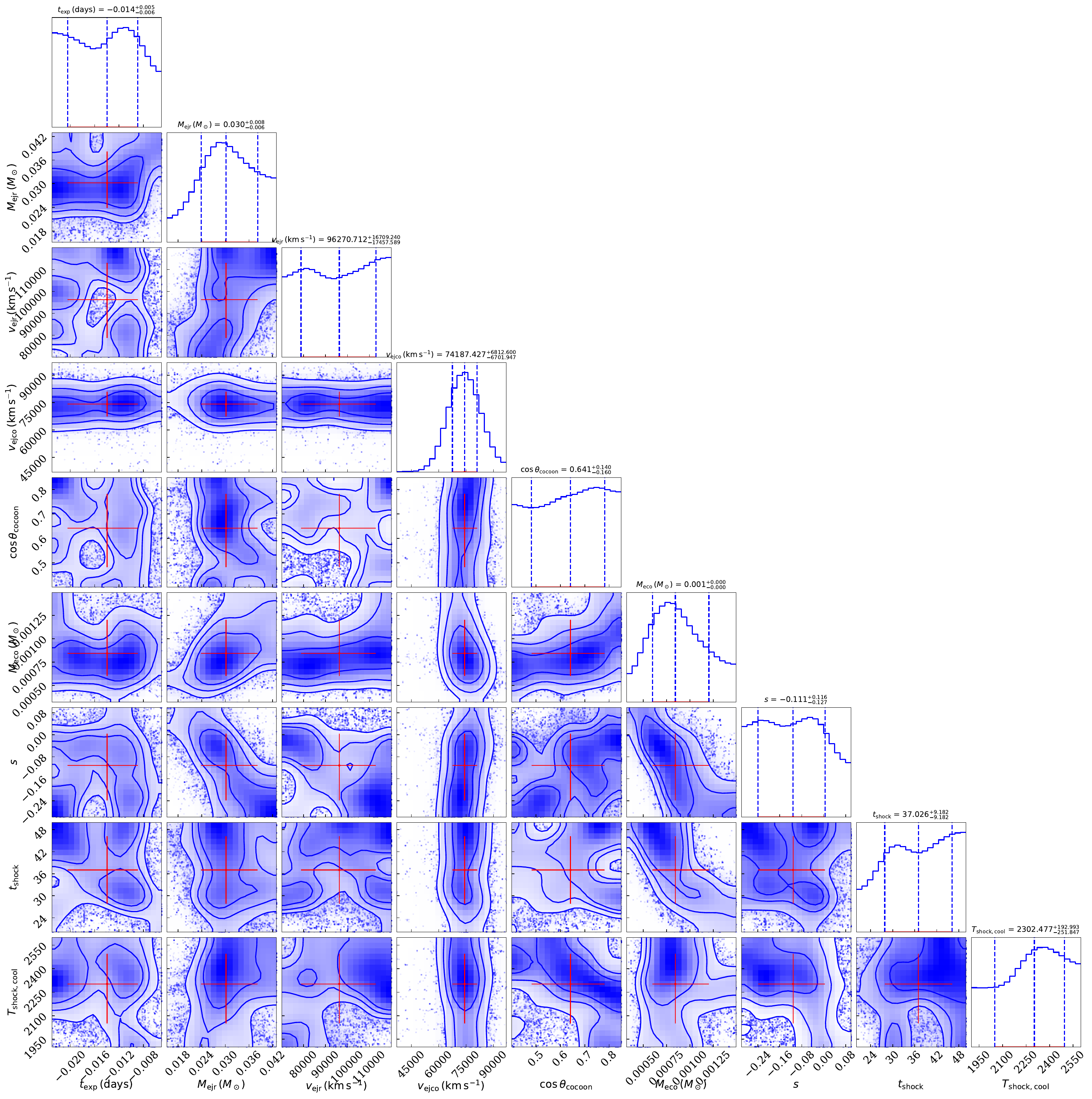} 
\caption{\textbf{Parameter constraints of the cocoon plus “red” kilonova model fitting for the afterglow-subtracted optical-NIR light curves of GRB 211211A (see Figure \ref{fig:co}).}  Histograms and contours show the likelihood map. Red crosses illustrate the best-fit values and 1-sigma error bars. The first parameter $t_{\text{exp}}$ donates the explosion time. The second and third parameters, $M_{\text{ejr}}$ and $v_{\text{ejr}}$, represent the mass and velocity of the “red” ejecta. The final six parameters describe the cocoon properties. $v_{\text{ejco}}$ and $M_{\text{eco}}$ are the velocity and mass of the cocoon (or blue) ejecta. 
$\cos \theta_{\text{cocoon}}$ represents the cocoon opening angle, and $s$ means the energy distribution of different velocity,
$dE/dv \propto v^{-s}$. $t_{\text{shock}}$ is the shock breakout timescale for the jet, and $T_{\text{shock, cool}}$ means the temperature floor of the cocoon.}
\label{fig:parco}
\end{figure*}

\begin{figure*}[tbp]
\centering\includegraphics[angle=0,height=6.2in]{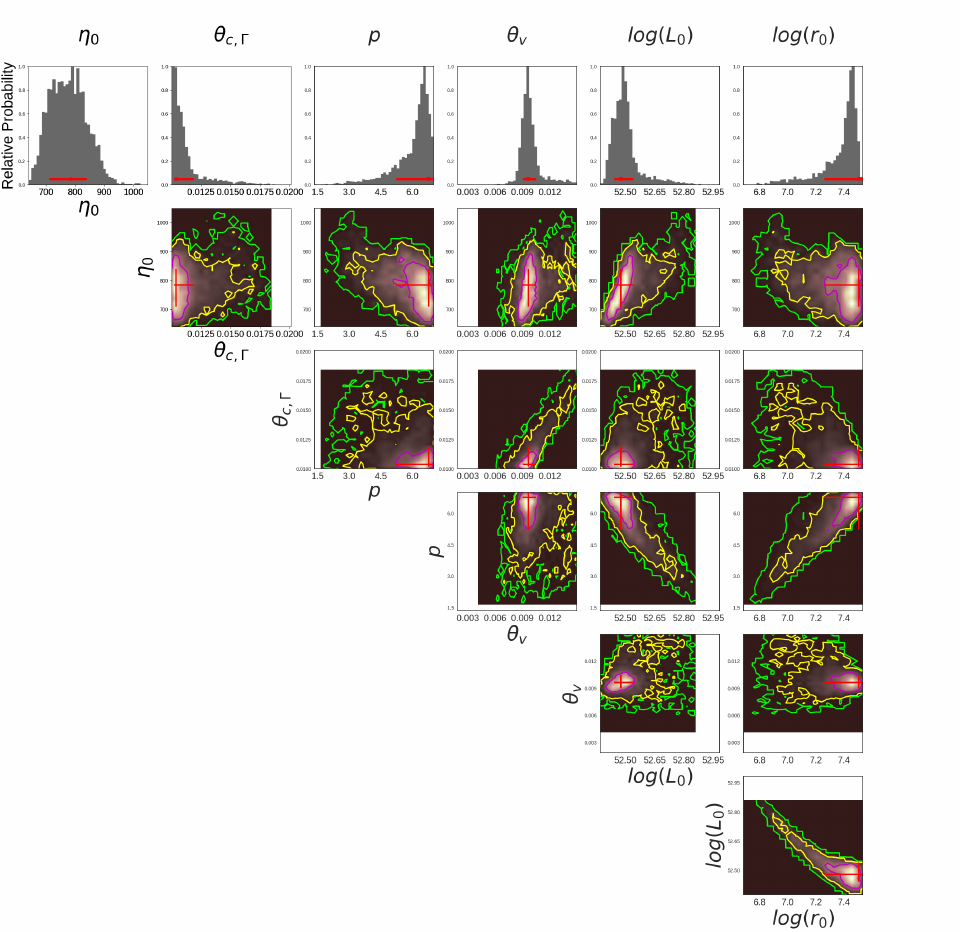} 
\caption{\textbf{Parameter constraints of our photosphere model fitting for the time-resolved spectrum measured from T0 + 2.0 s to T0 + 3.0 s.}  Histograms and contours show the likelihood map. Red crosses illustrate the best-fit values and 1-sigma error bars.}
\label{fig:par2}
\end{figure*}

\begin{figure*}[tbp]
\centering\includegraphics[angle=0,height=6.2in]{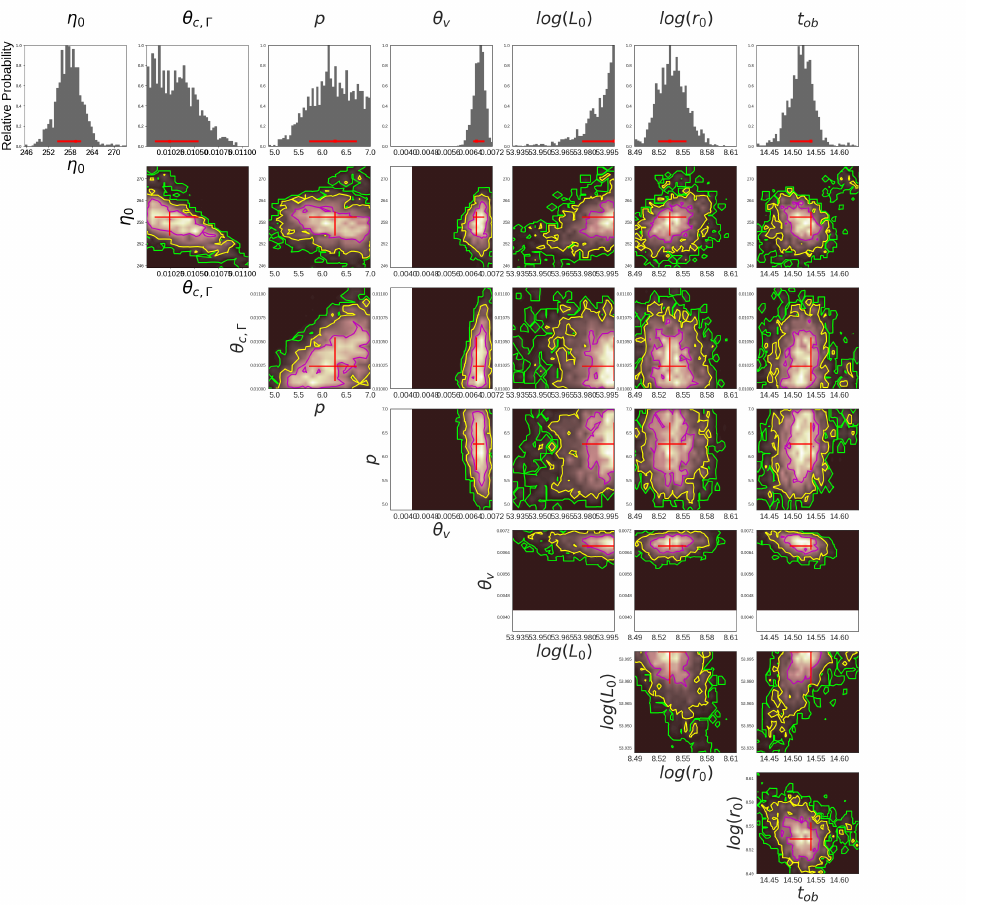} 
\caption{\textbf{Parameter constraints of our photosphere model fitting for the time-resolved spectrum measured from T0 + 12.0 s
to T0 + 14.0 s.} For the luminosity injection, $\tau _{1}=32$, $\tau_{2}=0.5$, $\hat{t}_{s}=3.4$ is adopted. The luminosity peaks at $\hat{t}_{p}=\hat{t}_{s}+\left( \tau _{1}\cdot \tau _{2}\right)^{1/2}=7.4$ s, corresponding to the second pulse.}
\label{fig:par12}
\end{figure*}

\begin{figure*}[tbp]
\centering\includegraphics[angle=0,height=6.8in]{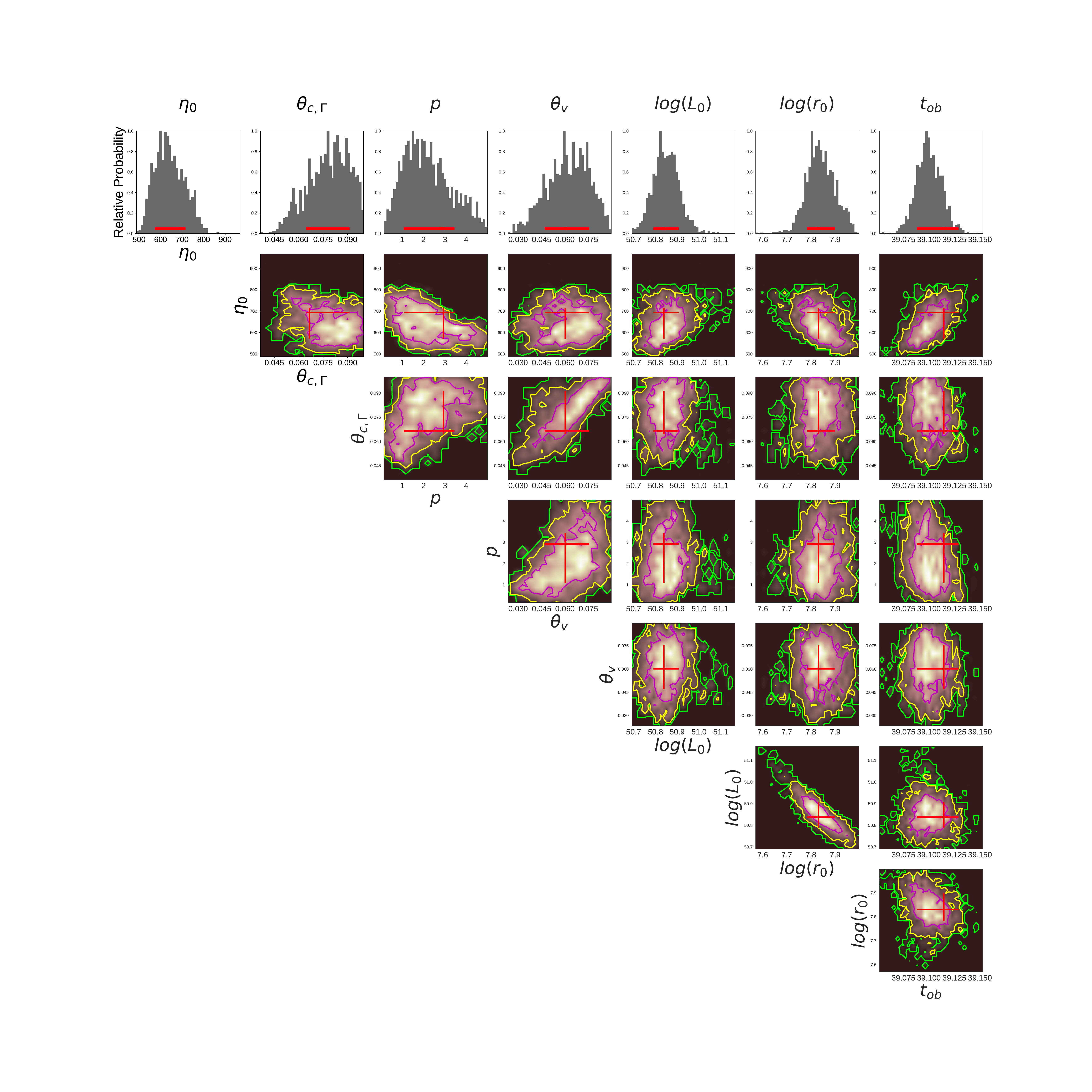} 
\caption{\textbf{Parameter constraints of our photosphere model fitting for the time-resolved spectrum measured from T0 + 40.0 s
to T0 + 45.0 s.} For the luminosity injection, $\tau _{1}=4$, $\tau_{2}=4$, $\hat{t}_{s}=20$ ($T_{90}$ $\sim$ 10 s) is adopted. The luminosity peaks at $\hat{t}_{p}=\hat{t}_{s}+\left( \tau _{1}\cdot \tau _{2}\right)^{1/2}=24$ s, corresponding to the late injection at $\sim$ 20 - 30 s.}
\label{fig:par40}
\end{figure*}

\begin{table}
\caption{\textbf{Spectral fitting parameters using the probability photosphere model with structured jet ($\theta _{c,\Gamma}<\theta _{c,L}$ and $\theta _{c,L}=0.1$), for 2.0 - 3.0 s, 12.0 - 14.0 s and 40.0 - 45.0 s in GRB 211211A. } Obviously, $\theta_{v} \gtrsim \theta_{c,\Gamma}$ ($\sim$ 1.0 - 1.5 times) is consistently obtained.}
\label{TABLE:MCpho}\center%
\begin{tabular}{lcccl}
\toprule Parameters & 2.0 - 3.0 s & 12.0 - 14.0 s & 40.0 - 45.0 s & \\
\midrule 
$\eta _{0}$ & $784.38_{-74.12}^{+55.24}$ &  $259.42_{-5.09}^{+1.52}$ &  $693.85_{-121.68}^{+21.36}$ & \\
$\theta_{c,\Gamma}$ (rad) & $0.01_{-0.00}^{+0.00}$ &  $0.01_{-0.00}^{+0.00}$ &  $0.06_{-0.00}^{+0.02}$ & \\
$p$ & $6.76_{-1.55}^{+0.24}$ & $6.26_{-0.54}^{+0.46}$ & $2.92_{-1.84}^{+0.53}$ & \\
$\theta _{\text{v}}$ (rad) & $0.01_{-0.00}^{+0.00}$ & $0.01_{-0.00}^{+0.00}$ & $0.06_{-0.01}^{+0.02}$ &  \\
log $L_{0}$ (erg s$^{-1})$ & $52.48_{-0.04}^{+0.06}$ & $53.99_{-0.02}^{+0.00}$ & $50.84_{-0.05}^{+0.07}$ & \\
log $r_{0}$ (cm) & $7.50_{-0.24}^{+0.03}$ &  $8.53_{-0.01}^{+0.02}$ &  $7.83_{-0.05}^{+0.07}$ &\\
$t_{\text{ob}}$ (s) &  &  $14.53_{-0.05}^{+0.00}$ &  $39.11_{-0.02}^{+0.02}$ &\\
\bottomrule &  &
\end{tabular}%
\end{table}

\end{document}